\documentclass[aps,pra,twocolumn,showpacs,floatfix,superscriptaddress]{revtex4-1}
\usepackage{color}      
\usepackage{amsmath,amssymb,amsfonts,bbm,bm,graphicx,hyperref,times,subfigure}
\usepackage[T1]{fontenc}

\renewcommand{\det}{{\rm Det}\,}

\newcommand{\be}{\begin{equation}}
\newcommand{\ee}{\end{equation}}
\newcommand{\bea}{\begin{eqnarray}}
\newcommand{\eea}{\end{eqnarray}}
\newcommand{\ket}[1]{|#1\rangle}
\newcommand{\bra}[1]{\langle#1|}

\def\Tr{\hbox{Tr}} \def\sigmaCM{\boldsymbol{\sigma}}
\begin{document}
\title{Quantum cooling and squeezing of a levitating nanosphere via time-continuous measurements}
\date{\today}
\author{Marco G. Genoni}
\affiliation{Department of Physics \& Astronomy, University College London, 
Gower Street, London WC1E 6BT, United Kingdom}
\email{marco.genoni@ucl.ac.uk}
\author{Jinglei Zhang}
\affiliation{Scuola Normale Superiore, I-56126 Pisa, Italy}
\author{James Millen}
\affiliation{Department of Physics \& Astronomy, University College London, 
Gower Street, London WC1E 6BT, United Kingdom}
\author{Peter F. Barker}
\affiliation{Department of Physics \& Astronomy, University College London, 
Gower Street, London WC1E 6BT, United Kingdom}
\author{Alessio Serafini}
\affiliation{Department of Physics \& Astronomy, University College London, 
Gower Street, London WC1E 6BT, United Kingdom}

\begin{abstract}
With the purpose of controlling the steady state of a dielectric nanosphere levitated within an optical cavity, we study its conditional dynamics under simultaneous sideband cooling and additional time-continuous measurement of either the output cavity mode or the nanosphere's position. We find that the average phonon number, purity and quantum squeezing of the steady-states can all be made more non-classical through the addition of time-continuous measurement. We predict that the continuous monitoring of the system, together with Markovian feedback, allows one to stabilize the dynamics for any value of the laser frequency driving the cavity. By considering state-of-the-art values of the experimental parameters, we prove that one can in principle obtain a non-classical (squeezed) steady-state with an average phonon number $n_{\sf ph}\approx 0.5$. 
\end{abstract}
\maketitle
\section{Introduction}

Bringing physical degrees of freedom to the quantum regime is proving so difficult that quantum control 
is bound to be a multi-branched endeavour, where techniques developed on different platforms and designed for different 
aims are blended together. 
In the context of cooling matter to the quantum ground state, a primary directive of quantum control, 
various techniques have come to the fore over the last twenty years. 
Prominent among them in the case of macroscopic mechanical systems is sideband cooling, where the targeted degree of freedom is driven by 
light on a red sideband, such that a beam-splitting light-matter interaction is achieved and excitations 
are drained out of the system, cooling it down. 
On the other hand, 
the implementation of efficient indirect quantum measurements is another 
obvious way to extract entropy from a quantum system. 
This study evaluates the combined performance of sideband laser cooling and continuous quantum measurements 
(also known as `monitoring') on a levitating nanosphere, an interesting opto-mechanical system where both 
such techniques are applicable to actual experiments.

The research field of quantum opto-mechanics, whose goal is to achieve control at the quantum level of massive mechanical oscillators, has received increasing attention in the last years, both for applicative and fundamental reasons \cite{RMPOptoMech}. 
So far, the main objective pursued by both theorists and experimentalists is the cooling of the oscillator either to its motional ground state or to non-classical states, such as low-number Fock states or squeezed states; several  protocols have been proposed in this respect, adapted to different physical settings and using different control strategies \cite{Clerk08,Genes08,Woolley08, Mari09, Hertzberg10,Cerillo10, OConnel10,Chan11,Teufel11,VannerPulsed,Verhagen12,Woolley13,Kronwald13,MauroNonClass, Szorkovsky2013,Suh14}.

Here, we will focus on a particularly promising setup, where the opto-mechanical system corresponds to a dielectric nanosphere trapped inside an optical cavity \cite{BarkerPRA10, Chang10, Kiesel13, Millen2014}. As the nanosphere is levitating, the coupling to the environment is minimised; it is possible then to neglect the thermal background of phonons which is typically one of the most detrimental sources of noise in opto-mechanical systems. Variations of this basic paradigm, where the nanosphere is trapped by the cavity field only \cite{Chang10, Monteiro2013}, by an optical tweezer within the cavity field \cite{Oriol2010}, or with the help of an electromagnetic trap \cite{Millen2014}, have been recently proposed. 
On the theoretical side, a detailed derivation of the master equation for the quantum state corresponding to the nanosphere's motion and to the cavity mode may be found in \cite{Oriol2011,Pflanzer2012}, and allows one to study the time behaviour of this opto-mechanical system, as well as its steady-state properties. As a matter of fact, this master equation paves the way to the analysis of protocols combining time-continuous measurement and feedback operations, and it will be the starting point of our study. 

We have witnessed constant progress in the understanding of quantum filtering, i.e. of the conditional dynamics of quantum systems subjected to time continuous measurements \cite{charmichael,WisemanMilburn,haus86,shapiro87,Belavkin,dalibard,wiseman93,wiseman94a,wiseman94b,doherty99,doherty00,steck04,ahn02,gammelmark,pinja,serafozzireview}. In particular, as regards systems described by continuous, canonical degrees of freedom, diffusive dynamics described by multivariate Wiener increments have been characterised in detail, and a general framework is available \cite{WisemanDiosi,WisemanDoherty}. 
Such diffusive dynamics correspond to conditional evolutions due to the monitoring of the environment through the class of so called {\em general-dyne} detections \cite{WisemanMilburn}. These quantum measurements amount to performing homodyne detections on the environment and, possibly, additional ancillary modes which are coupled to the environment itself via Gaussian unitary transformations \cite{ChiaWise,GeneralDino}. We recall that the term ``homodyne'' detection refers to the projective measurement on the eigenbases of canonical position and momentum operators $\hat{x}$ and $\hat{p}$. In the case where the overall Hamiltonian is quadratic in the canonical operators, and the system is linearly coupled to the environment, the conditional dynamics due to general-dyne detections preserves the Gaussianity of the quantum state and thus the whole dynamics can be equivalently described by the evolution of first and second statistical moments 
only. In this case, it is easy to optimise different steady-state properties, such as entanglement, squeezing and purity, over the parameters characterising the detection scheme \cite{mancini06,mancini07,serafozzi10,nurdin12,boundTH,Ravotto}.

The role of conditional dynamics due to indirect measurements, and of subsequent quantum feedback, for steering the quantum state of a mechanical oscillator towards either its ground state or a certain non-classical state has been already discussed in the literature. In \cite{DohertyJacobs1999}, Doherty and Jacobs derive the effective stochastic master equation corresponding to position measurements obtained through adiabatic elimination of a cavity mode continuously monitored through homodyne detection, and discuss the related feedback strategies aimed cooling the motion of the oscillator. 
The possibility to observe mechanical squeezing via a continuous {\em back-action evasion} measurement was firstly proposed in \cite{Clerk08} and then invesitgated experimentally in \cite{Hertzberg10,Suh14}. 
In \cite{DohertyPhylTrans2012}, the unravelling corresponding to direct position measurements of an oscillator interacting with a non-zero temperature thermal bath is considered; more specifically, the results obtainable in different regimes, corresponding to different measurement resolutions, are discussed in great detail, and the ensuing stochastic master equation has been seminal in the design of protocols to engineer thermo-mechanical squeezing \cite{Szorkovsky2011,Szorkovsky2013}. Also, the usefulness of discrete and repeated measurements on a coupled qubit in a hybrid setup has been investigated in \cite{ConMauro2015}, where an effective dynamics able to prepare a squeezed steady-state has been identified.
Very recently, Hofer and Hammerer \cite{Hofer2014} have studied the effect of continuous homodyne detection on the cavity output combined with sideband cooling in a standard opto-mechanical setup, where the mechanical oscillator interacts with a non-zero temperature thermal bath. First, they consider a single oscillator and discuss the corresponding steady-state average number of phonons; then, they present more complex and sophisticated protocols able, for example, to create entanglement between two distant oscillators.

As already stated above, in this manuscript we focus on the case of a levitating dielectric nanosphere in an optical cavity, described by the master equation derived in \cite{Pflanzer2012}. Notice that our treatment is distinct from the existing literature 
in that the master equation of the levitating nanosphere includes a photon scattering term, and the
measurements of both cavity output, via homodyne detection, and oscillator position, through the light scattered by the nanosphere itself, are considered simultaneously. By combining these measurements with sideband cooling and Markovian feedback, we address the possibility of both cooling the oscillator towards its ground state and of generating quantum mechanical squeezing, that is sub-vacuum fluctuations, which is a paradigmatic signature of non-classicality useful for quantum metrology and precision sensing \cite{Squeezing}. Finally, we also address in more detail the experimental setup described in \cite{Monteiro2013}, where the nanosphere is trapped in a high finesse optical cavity: by considering state-of-the-art values for the experimental parameters and for the measurement efficiencies, we show the possibility to vastly improve the performances of sideband cooling both in terms of steady-state average number of phonons and in terms of generation of squeezed quantum states. 

The manuscript is organised as follows: in Sec. \ref{s:ME} we discuss the master equation of a levitating nanosphere, and introduce the notation and figures of merit that will be discussed in the remainder of the article. In Sec. \ref{s:SME} we introduce the stochastic master equation describing the time-continuous measurements and then present the results obtainable for different values of their measurement efficiencies. In Sec. \ref{s:bead} we discuss the performances of these protocols for a specific experimental setup, while Sec. \ref{s:conclusions} concludes the paper with some final remarks.

\section{Levitating dielectric nanosphere master equation} \label{s:ME}

We will consider two quantum degrees of freedom; the cavity electromagnetic mode and the mechanical motion of a trapped nanosphere, described respectively by bosonic operators $a$ and $b$ satisfying the commutation relations $[a,a^\dag]=[b,b^\dag]=1$. We can then define the corresponding position and momentum quadrature operators as $x_c = (a+a^\dag)/\sqrt{2}$, $p_c = -i (a-a^\dag)/\sqrt{2}$, $x_m = (b+b^\dag)/\sqrt{2}$ and $p_m = -i (b-b^\dag)/\sqrt{2}$, which can be grouped in a single vector 
\begin{align}
{\bf r} = (x_c,p_c, x_m,p_m)^{\sf T}. \label{eq:quadrvector}
\end{align} 
By considering the cavity driven by a laser at frequency $\omega_L$, the Hamiltonian describing the interaction between the two modes reads
\begin{align}
H = \omega_m b^\dag b - \Delta a^\dag a + g (a+a^\dag)(b+b^\dag), \label{eq:Hamilton}
\end{align}
where $g$ is the effective coupling constant, $\omega_m$ is the mechanical frequency and we have already transformed the Hamiltonian to a frame rotating at the driving laser frequency $\omega_L$, such that $\Delta=\omega_L - \omega_c$ denotes the detuning from the cavity resonance $\omega_c$ (note that we set $\hbar=1$). By considering the open dynamics resulting from the interaction with the environment ({\em i.e.} the {\em free} electromagnetic modes), one obtains the master equation \cite{Pflanzer2012},
\begin{align}
\frac{d\varrho}{dt} &= \mathcal{L}\varrho \nonumber \\
&= - i [H,\varrho] + \kappa \: \mathcal{D}[a]\varrho + \Gamma \: \mathcal{D}[b+b^\dag]\varrho \:,
\label{eq:ME}
\end{align}
where $\mathcal{D}[O]\varrho=O\varrho O^\dag - (O^\dag O \varrho + \varrho O^\dag O)/2$. The first term is responsible for the unitary dynamics, the second one describes the usual cavity loss (with total loss rate $\kappa$), while the third one corresponds to the recoil heating due to photon scattering from the oscillating nanosphere (with decoherence rate $\Gamma$). The dependence and formulas for all the parameters entering in Eqs. (\ref{eq:Hamilton}) and (\ref{eq:ME}) can be found in \cite{Pflanzer2012}. 
Specifically we want to point out that the cavity loss parameter $\kappa=\kappa_0 + \kappa_d$ is the sum of the intrinsic loss rate $\kappa_0$ due to the imperfections in the cavity mirrors, plus the extra contribution $\kappa_d$ due to the presence of the dielectric inside the cavity. We also remark that we only address the control along one spatial direction (dictated by the harmonic trap generated by the optical tweezers) along the optical cavity axis and we will not deal with the potential technicalities involved in cooling the motion along the other two decoupled directions.
\\

By assuming that the system is prepared in a Gaussian state (e.g. in a thermal state) at time $t=0$, at every time the dynamics keeps the state Gaussian (see \cite{GaussAOP} for different reviews on Gaussian states). As a consequence, the whole dynamics can be fully described by means of the first moment vector ${\bf R}$ and the covariance matrix $\sigmaCM$, whose elements are defined as
\begin{align}
R_j &= \Tr[\varrho r_j] \nonumber \\
\sigma_{jk} &= \Tr[\varrho (r_j r_k + r_k r_j)] - 2 R_j R_k . \label{eq:CM}
\end{align}
Throughout the article, we will discuss the efficiency of our protocols by considering the effect on the mechanical oscillator properties. Hence, it is useful to introduce the covariance (sub)matrix corresponding to the oscillator quantum state alone, obtained by tracing out the cavity mode, $\varrho^{(m)} = \Tr_c [\varrho]$. This corresponds to
\begin{align}
\sigmaCM^{(m)} =
\left(
\begin{array}{c c}
\sigma_{33} & \sigma_{34} \\
\sigma_{34} & \sigma_{44}
\end{array}
\right)
= 2 \left(
\begin{array}{c c}
\Delta x_m^2 & \Delta x_m p_m \\
\Delta x_m p_m & \Delta p_m^2
\end{array}
\right) \:,
\end{align}
where $\sigma_{jk}$ are the elements of the global covariance matrix defined in Eq. (\ref{eq:CM}), which in fact correspond to variances and covariances of the quadrature operators $x_m$ and $p_m$. 
All the properties we are interested in can be easily obtained from the matrix $\sigmaCM^{(m)}$. We will focus on three figures of merit: quantum squeezing, purity and average number of phonons. 

Quantum squeezing can be quantified through the minimum eigenvalue of $\sigmaCM^{(m)}$, as 
\begin{align}
\xi = \min \textrm{eig}[\sigmaCM^{(m)}].
\end{align}
In the next sections we will show the squeezing behaviour, by plotting $\xi$ in dB scale, such that negative values will correspond to a quadrature of the mechanical oscillator having sub-vacuum fluctuations, and thus to a non-classical squeezed state. In general this quadrature will be a certain linear combination of position and momentum operators, and the detectability and usefulness of the corresponding squeezing may not be straightforward. This is one of the reasons why, in Sec. \ref{s:bead}, we will also focus on the position fluctuations $\Delta x_m^2$ only. 

The usefulness of a certain quantum state in quantum information and quantum communication protocols often strictly depends on its purity, that is 
on how close a state $\varrho$ is to a projector $\ket{\psi}\bra{\psi}$ on a single Hilbert space vector, rather than a statistical mixture thereof. 
Such single vector states are also known as pure states. 
The purity of a quantum state is defined as $\mu= \Tr[\varrho^2]$ and it takes is maximum value $\mu=1$ if and only if $\varrho$ is a pure state. For Gaussian states, the purity can be evaluated through the covariance matrix, and in particular for the single-mode mechanical oscillator we can use the formula $\mu = 1/\sqrt{\det[\sigmaCM^{(m)}]}$. Notice that, as we are considering single-mode states, all entropies, including the von Neumann entropy, are monotonic functions of the purity, which thus fully characterize the mixedness of the quantum state.

Finally, as we will mainly consider steady-states of the mechanical oscillator having zero first moments ($\Tr[\varrho x_m]=\Tr[\varrho p_m]=0$), 
the number of phonons can also be evaluated directly from the covariance matrix as
\begin{align}
n_{\sf ph} &= \Tr[\varrho b^\dag b] =\frac{\Tr[\varrho (x_m^2 + p_m^2)]-1}{2} \nonumber \\
&= \frac{\hbox{tr}[\sigmaCM^{(m)}]-2}{4} . \label{eq:nphon}
\end{align}
Here, $\hbox{tr}[\cdot]$ denotes to the trace of a finite dimensional matrix. Notice that whenever the first moments of the oscillator are not equal to zero, Eq. (\ref{eq:nphon}) gives only a lower bound on the actual average number of phonons $n_{\sf ph}$. 

In terms of first moment vector and of the global covariance matrix, the master equation for the quantum state $\varrho$ in (\ref{eq:ME}) entails the 
following time evolution:
\begin{align}
\frac{d {\bf R}}{dt} &= A {\bf R} \, , \\
\frac{d \sigmaCM}{dt} &= A \sigmaCM + \sigmaCM A^{\sf T} + D \, , 
\end{align}
where the matrices $A$ (drift matrix) and $D$ (diffusion matrix) can be easily evaluated from the parameters entering in (\ref{eq:ME}) and in the interaction Hamiltonian (\ref{eq:Hamilton}) \cite{WisemanDoherty} and are reported in the Appendix \ref{s:appendix}. The existence of a steady-state of the dynamics (having zero first moments) can be easily discussed by analyzing the eigenvalues of the drift matrix $A$:  
\begin{align}
\textrm{stable dynamics} \:\: \Leftrightarrow \:\: \hbox{Re}[\alpha_j] <0  \qquad \forall j
\end{align}
where $\hbox{Re}[x]$ denotes the real part of a complex number $x$, and $\alpha_j$ are the eigenvalues of the drift matrix $A$.
Such a property is often referred to as `Hurwitz stability' in the control literature.

\begin{figure}[h]
\begin{center}
\includegraphics[width=0.9\columnwidth]{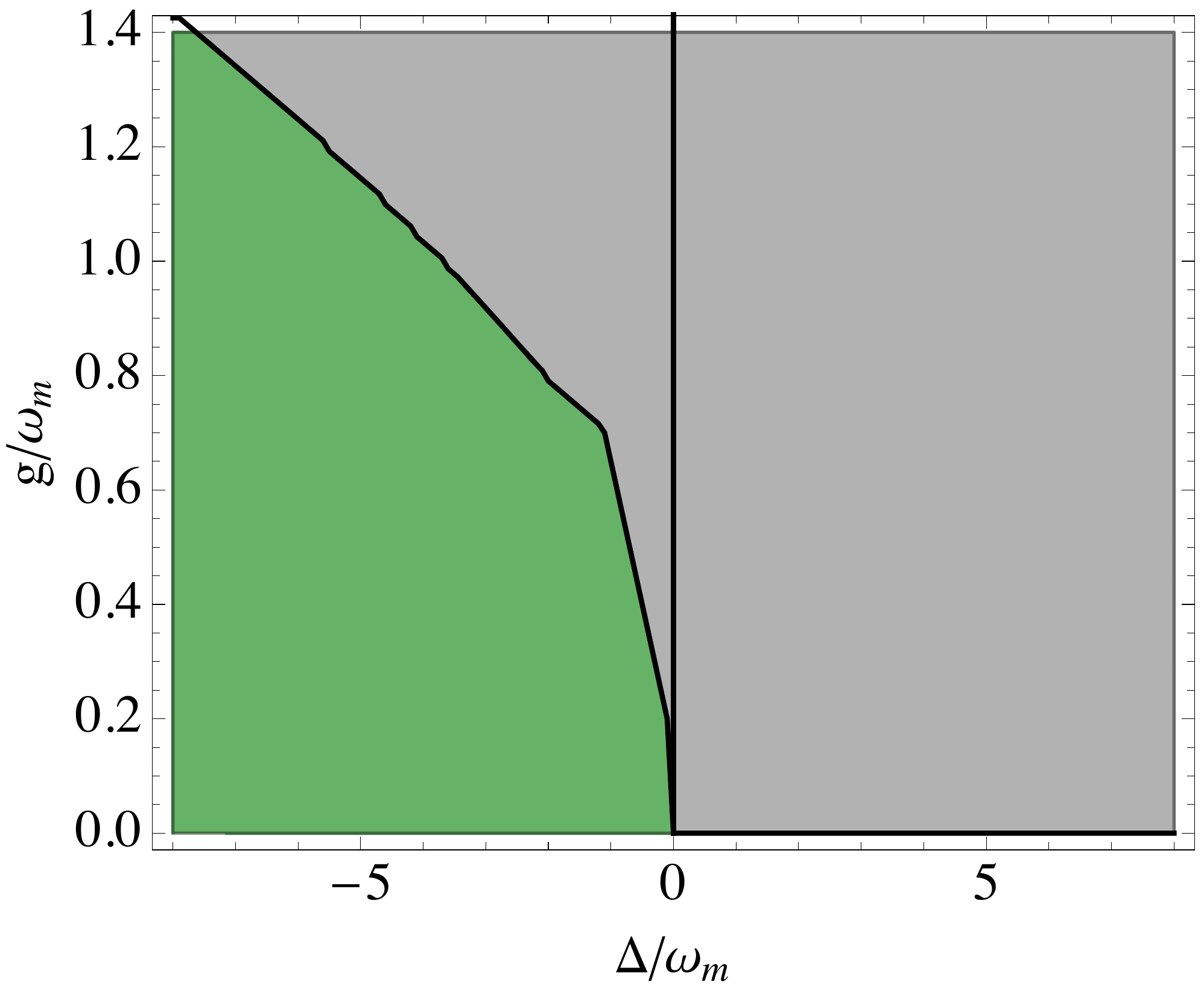}
\end{center}
\caption{Stable (green) and unstable (gray) regions for the master equation (\ref{eq:ME}), as a function of the laser detuning $\Delta$ and of the oscillator-cavity mode coupling constant $g$, for $\kappa=2\omega_m$ and $\Gamma=\omega_m/10$. Notice that, for $\Delta/\omega_m =0$, the drift matrix $A$ always has an eigenvalue equal to zero, and thus the system cannot be considered strictly stable.
\label{f:stability}}
\end{figure}

\begin{figure}[h]
\begin{center}
\includegraphics[width=0.48\columnwidth]{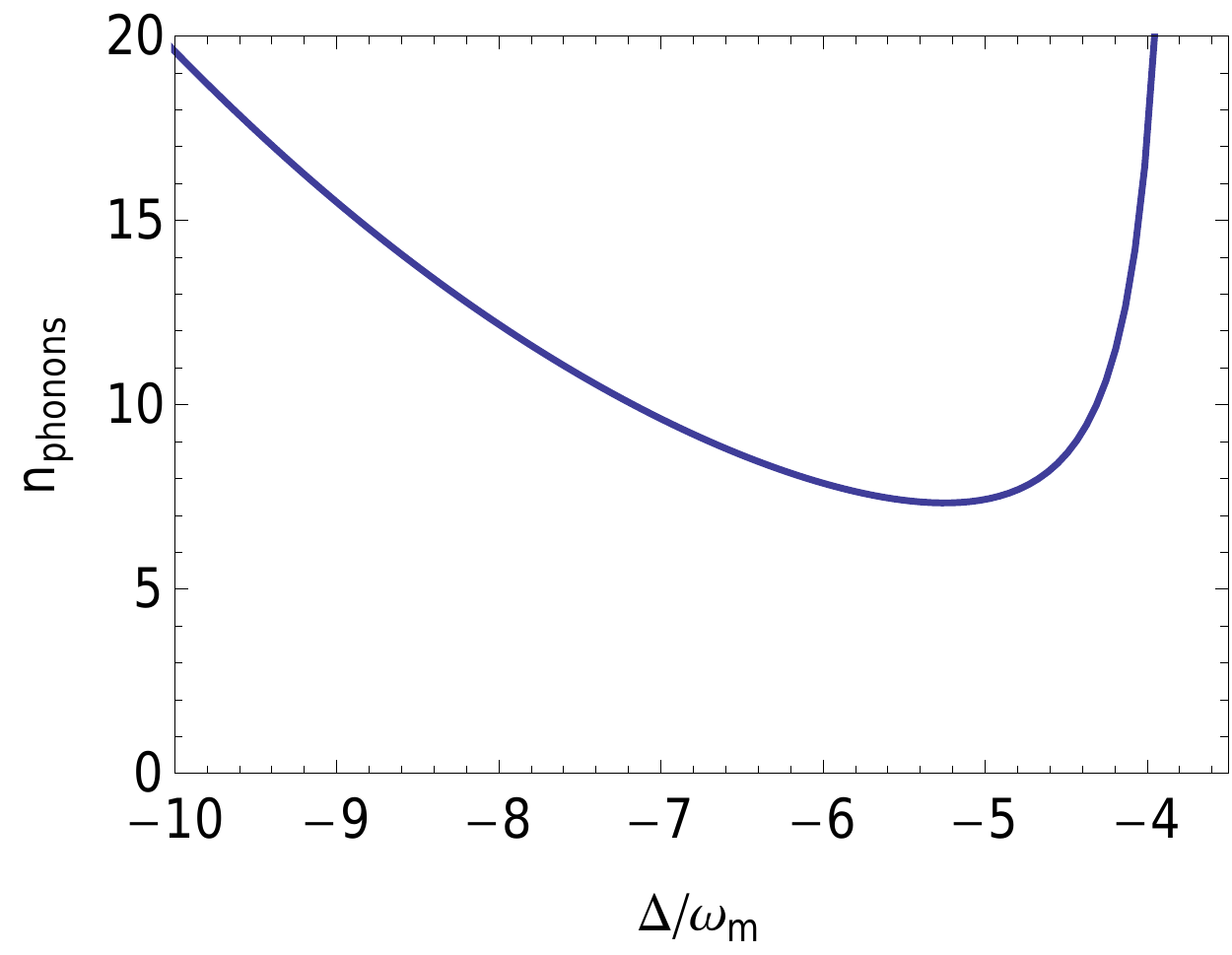}\medskip \
\includegraphics[width=0.48\columnwidth]{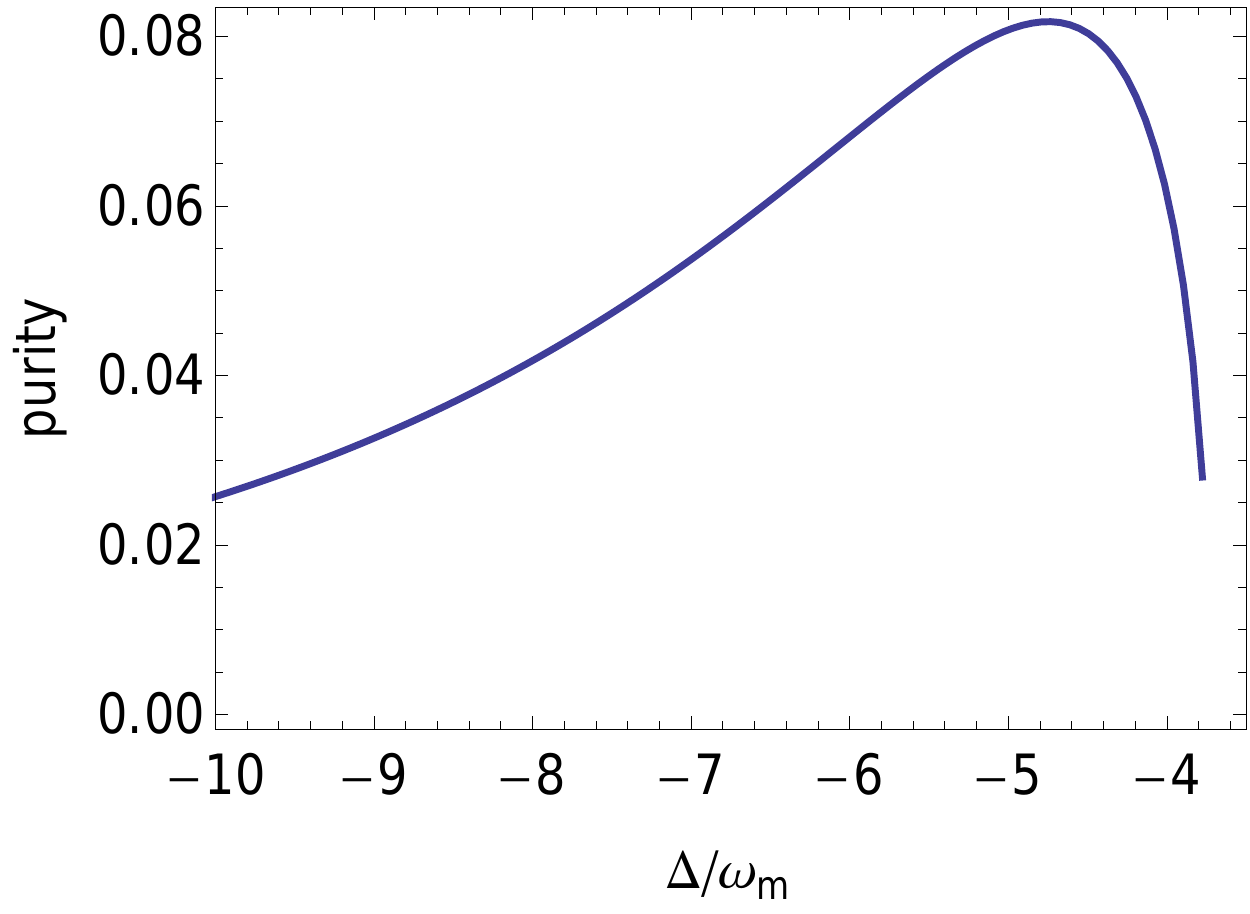}
\end{center}
\caption{
Steady-state values for average number of phonons (left) and for the oscillator purity (right), as a function of the detuning $\Delta$ ($g=\omega_m$, $\kappa=2\omega_m$ and $\Gamma=\omega_m/10$). Notice that we plot only values of detuning corresponding to the stable region and that minimum of phonon number and maximum of purity do not exactly correspond (this is due to the fact that the steady-state corresponds to a squeezed thermal state).
\label{f:nomeas}}
\end{figure}

A plot of the stable regions for the master equation (\ref{eq:ME}) as a function of the detuning $\Delta$ and of the coupling constant $g$ is pictured in Fig. \ref{f:stability}. 
As one can see, stability is obtained only for red-detuning ($\Delta<0$) and, in order to reach a steady-state, for larger values of the coupling constant $g$, one needs also a larger value of the detuning $ | \Delta|$. 
In Fig. \ref{f:nomeas}, we plot the values of the purity of the oscillator and the average number of phonons obtainable at steady-state as a function of the detuning $\Delta$, where the other parameters 
have been given plausible experimental values in current setups as in \cite{Pflanzer2012}; one can see that for large red-detunings one can  cool the oscillator to a state with around $n_{\sf ph}=8$ phonons at steady-state. \\
\section{Time-continuous homodyne measurement of cavity mode and oscillator position} \label{s:SME}
We now consider the conditional evolution due to general-dyne time-continuous measurement on both the cavity mode and the oscillator. The corresponding stochastic master equation reads
\begin{align}
d\varrho = \mathcal{L}\varrho \: dt + \sqrt{\eta_1 \kappa} \: \mathcal{H}[a e^{i \phi}] \varrho \: dw_1 + \sqrt{\eta_2 \Gamma} \: \mathcal{H}[b+b^\dag] \varrho \: dw_2 
\label{eq:SME}
\end{align}
where $\mathcal{H}[O]\varrho = O\varrho + \varrho O^\dag - \Tr[(O+O^\dag)\varrho]\varrho$ and $dw_j$ are uncorrelated Wiener increments, such that $dw_j dw_k = dt \: \delta_{jk}$. The term $\sqrt{\eta_1 \kappa}\mathcal{H}[a e^{i\phi}]$ describes the effect of continuous homodyne on the output cavity mode with efficiency $\eta_1$, where the phase $\phi$ can be adjusted by choosing the optical phase of the monitored quadrature operator ({\em e.g.} $\phi=0$ and $\phi=\pi/2$ correspond respectively to homodyning quadratures $x_c$ and $p_c$) \cite{WisemanMilburn,WisemanDoherty}.
Analogously, the term  $\sqrt{\eta_2 \Gamma} \: \mathcal{H}[b+b^\dag]$ describes the effect of continuous monitoring of the oscillator position, with efficiency $\eta_2$ \cite{DohertyJacobs1999,DohertyPhylTrans2012}.

As for the unconditional master equation (\ref{eq:ME}), the dynamics induced by the continuous measurement here considered does not change the Gaussian character of the quantum state; as a consequence we can translate Eq.~(\ref{eq:SME}) into equations for the first moment vector and covariance matrix:
\begin{align}
d {\bf R} &= A {\bf R} \:dt + (N - \sigmaCM B^{\sf T}) \:d{\bf w} \, , \label{eq:firstmoment} \\
\frac{d \sigmaCM}{dt} &= \widetilde{A} \sigmaCM + \sigmaCM {\widetilde{A}}^{\sf T} -\sigmaCM B^{\sf T} B \sigmaCM + \widetilde{D} \:,   \label{eq:Riccati}
\end{align}
where $d{\bf w} = (dw_1, dw_2)^{\sf T}$ and the matrices  $N, B, \widetilde{A}$ and $\widetilde{D}$ can be evaluated starting from the parameters entering the stochastic master equation (\ref{eq:SME}) \cite{WisemanDoherty} and are reported in the Appendix \ref{s:appendix}.
It is important to observe that the Riccati equation for the covariance matrix is completely deterministic, and yields a steady-state that can be efficiently evaluated numerically. On the other hand, the first moments' evolution is stochastic, i.e.~it depends on the outcomes of the continuous measurements. As a consequence, at each time the conditional state is a Gaussian state whose covariances and correlations evolve deterministically according to the dissipative dynamics and the kind of measurement performed, while its first moments evolves randomly in the phase-space, depending on the values of the photocurrents.
In the following we will focus on these conditional steady-states only.

Although it is possible to achieve these conditional covariance matrices by pure filtering, i.e.~recording the measurement outcomes (photocurrents), in order to remain in the harmonic trap regime, where our treatment applies, 
it is useful to suppress the drift of the first moments, due to the stochastic evolution, by an active feedback operation. The role of feedback is indeed to use the information contained in the measurement outcomes in order to remove the contribution given by the last term in Eq. (\ref{eq:firstmoment}), which is proportional to the Wiener increment $d{\bf w}$.  This can always be done by adding a linear feedback term in the Hamiltonian (\ref{eq:ME}) with coupling constants proportional to the photocurrents, i.e.
\begin{align}
H^\prime  = H + {\bf r}^{\sf T} {\bf f}(t) \; ,
\end{align}
where ${\bf r}$ is the vector of quadrature operators introduced in Eq. (\ref{eq:quadrvector}) and ${\bf f}(t)$ is an optimized vector of time-dependent coupling constants whose values depends linearly on the continuous-measurement outcomes \cite{WisemanDoherty}.
In practice, while for the cavity field this corresponds simply to a linear driving, in the case of a mechanical oscillator it can be obtained by means of a combination of impulses and shifts of the trapping potential (for a more detailed discussion of this issue see \cite{DohertyJacobs1999}).

The first important consequence of Eqs. (\ref{eq:SME}) and (\ref{eq:Riccati}) regards the stability of the opto-mechanical system.
The existence of a steady-state for a continuously monitored quantum systems has been discussed in \cite{WisemanDoherty}. More specifically, it is proven that Eq. (\ref{eq:Riccati}) has a stabilizing solution if and only if the pair of matrices $(B,\widetilde{A})$ is {\em detectable}, namely
\begin{align}
B {\bf x}_\lambda \neq 0 \:\:\:\: \forall \: {\bf x}_\lambda : \widetilde{A}{\bf x}_\lambda = \lambda {\bf x}_\lambda \:\: \textrm{with} \: {\rm Re}[\lambda] \geq 0 \:, 
\end{align}
that is whenever the degrees of freedom that are not strictly stable under the drift matrix $\widetilde{A}$ contribute to the measurement output $B {\bf r}$. We find that, for all the choices of parameters we have considered in our numerical simulation, whenever the interaction between the two bosonic modes is on (i.e. for $g>0$), if a continuous measurement is performed, {\em i.e.} if $\eta_1>0$ or $\eta_2>0$, the stochastic master equation satisfies the stability conditions.
We should remark that this stability condition regards the covariance matrix steady-state, while in principle the first moments could not go to a steady-state value (e.g. to zero). However, as we have just stated above and discussed for example in \cite{DohertyJacobs1999}, the information obtained from the measurement can be used to obtain a proper steady-state for the quantum system with zero first moments, as the stochastic drift 
on the latter may always be canceled by Markovian linear feedback.

In the following, we will concentrate on the the steady-state properties of the harmonic oscilaltor. As anticipated in the previous section, we will 
analyse the number of phonons, the purity of the state, and at the achievable quantum squeezing, quantified by the minimum eigenvalue of the steady-state covariance matrix. We will consider different measurement strategies: (i) measurement of the cavity mode only ($\eta_1>0$ and $\eta_2=0$); (ii) measurement of the oscillator position only ($\eta_1=0$ and $\eta_2>0$); (iii) simultaneous measurement of the cavity mode and of the oscillator ($\eta_1>0$ and $\eta_2>0$).

\subsection{Time-continuous measurement of the cavity mode} 
\begin{figure}[t]
\begin{center}
\includegraphics[width=0.8\columnwidth]{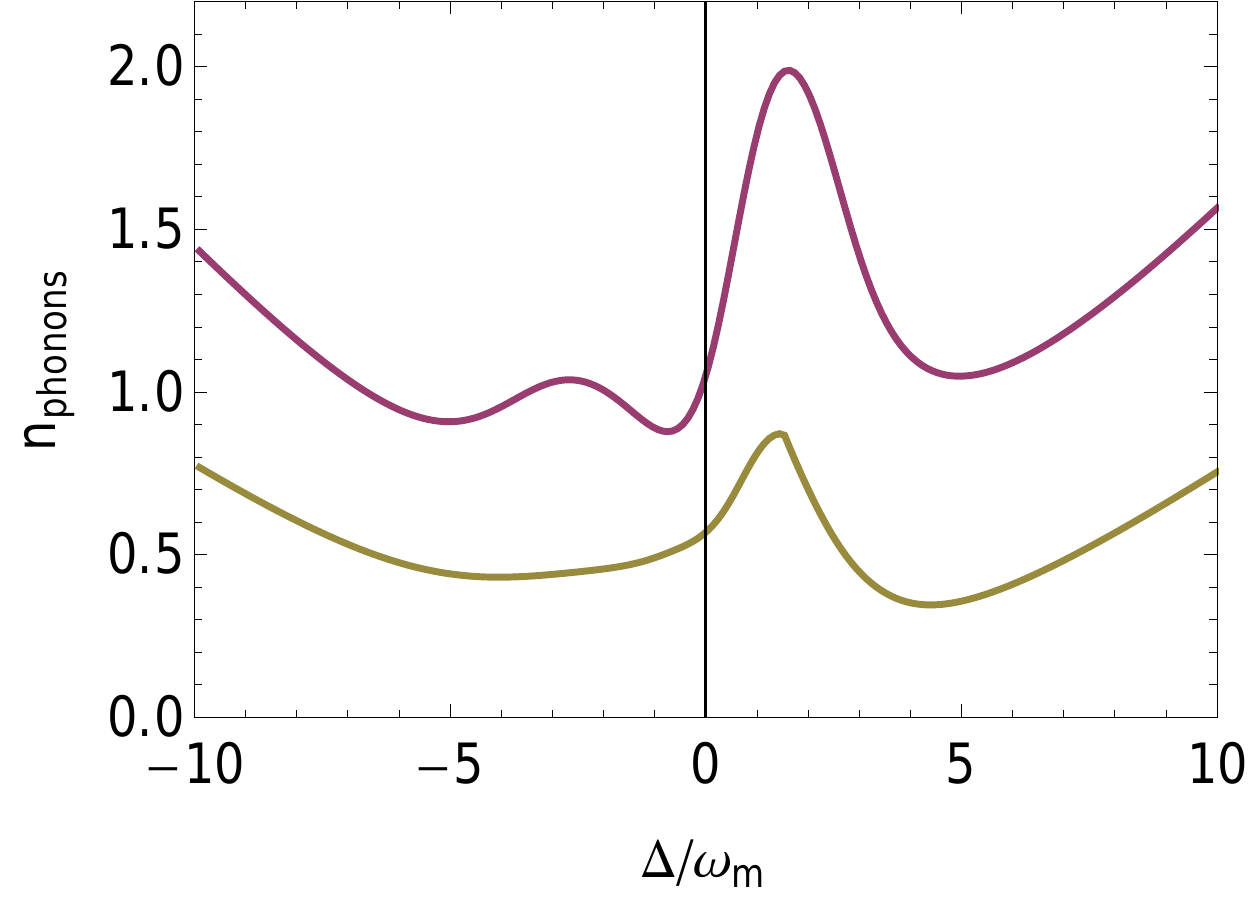}\ \\
\includegraphics[width=0.48\columnwidth]{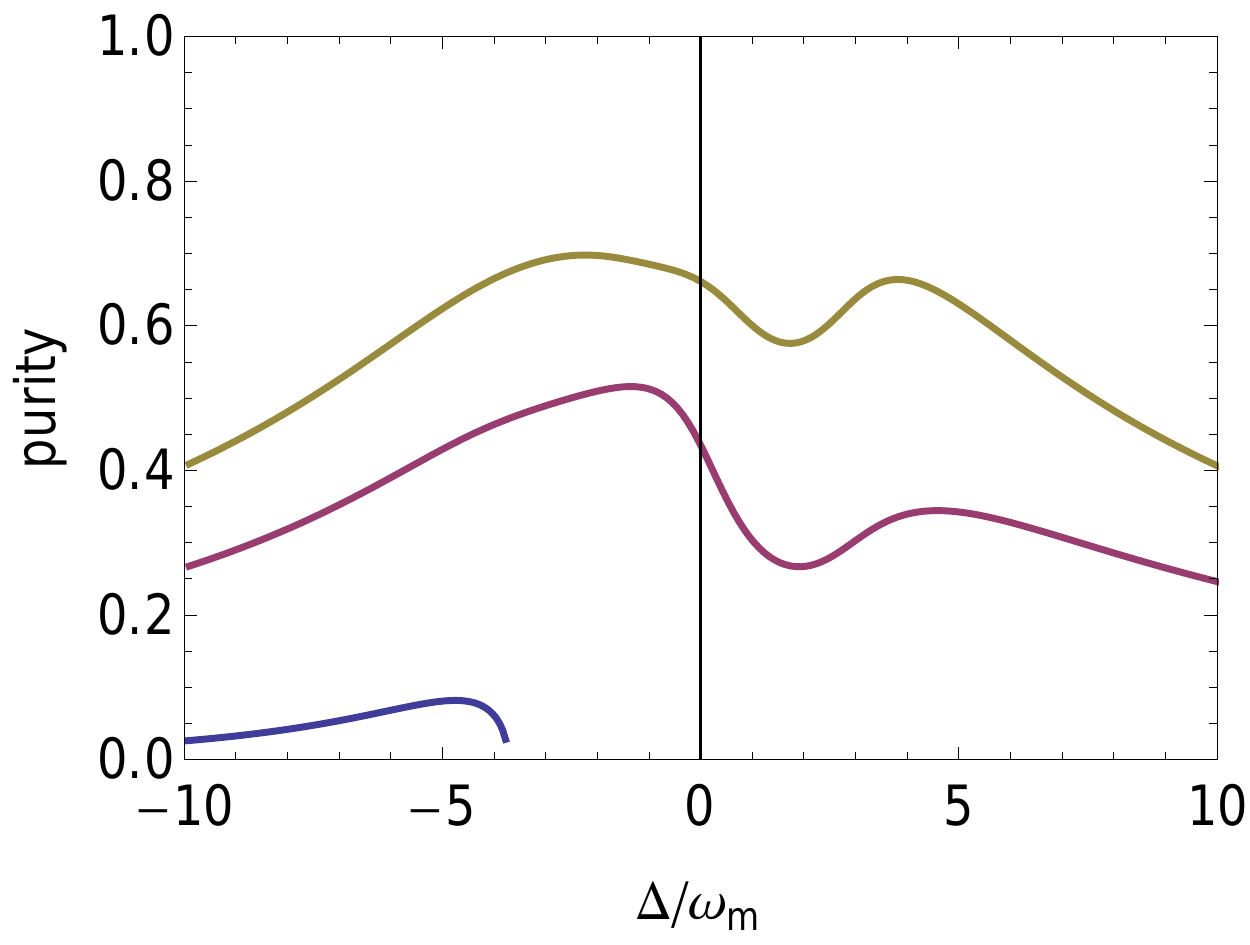} \medskip \
\includegraphics[width=0.48\columnwidth]{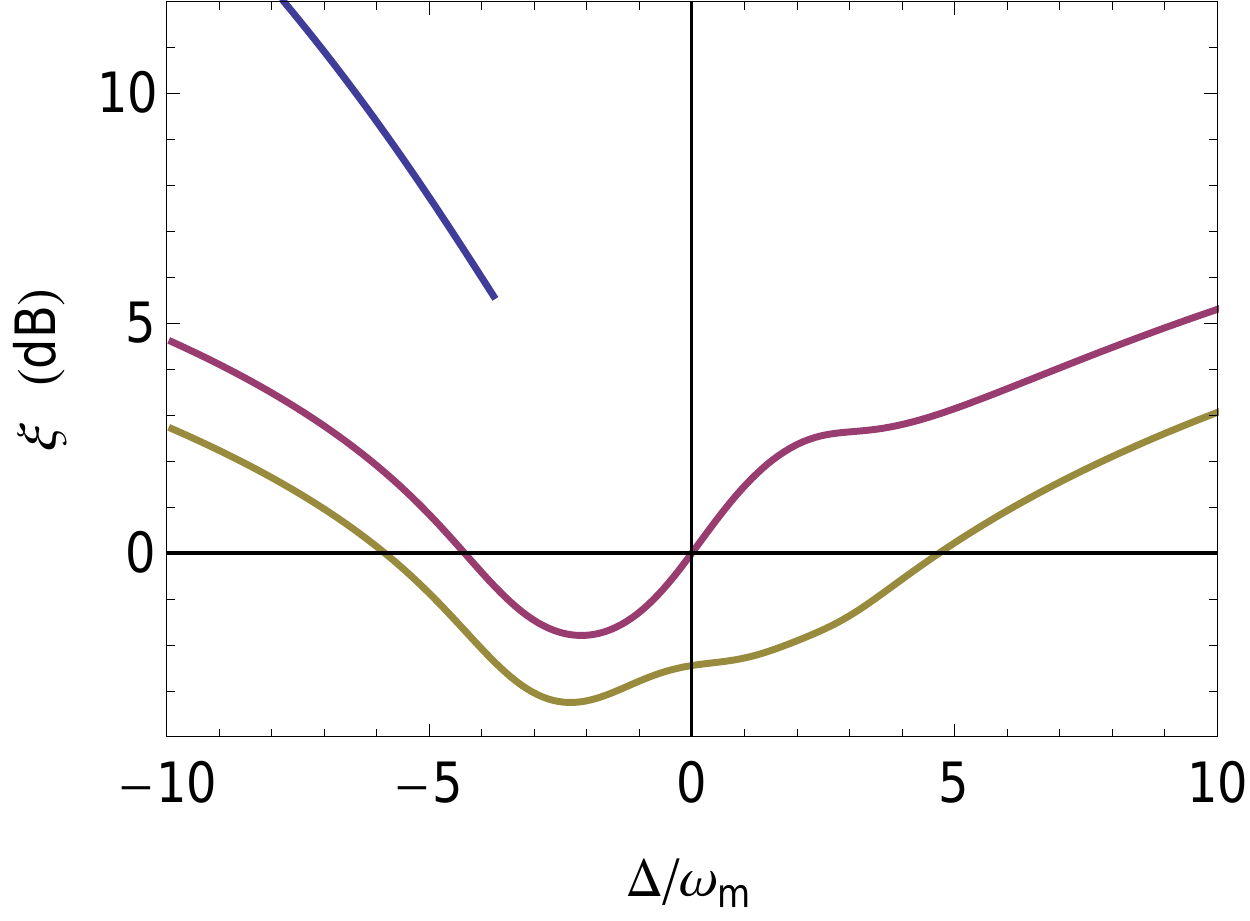}
\end{center}
\caption{Results obtained via continuous optimized homodyne of the output cavity mode. Top: Steady-state values for average number of phonons. Bottom: steady-state oscillator purity (left) and quantum squeezing in dB scale (right). All quantities are plotted as a function of the detuning $\Delta$ and for different values of the cavity mode measurement efficiency: blue, $\eta_1=0$ (notice that the blue lines are plotted only in the squeezing and purity plots, and only for the small region of values where the system is stable); purple, $\eta_1=0.4$; yellow, $\eta_1=1$.
The other parameters are fixed as follows: $\eta_2=0$, $g=\omega_m$, $\kappa=2 \omega_m$ and $\Gamma=\omega_m/10$.
\label{f:cavitymeasP}}
\end{figure}

In this subsection we investigate the properties of the steady-state of the oscillator in the case where no measurement is performed directly on the nanosphere, while the output of the cavity is continuously measured. 
The phase $\phi$ of the quadrature which is monitored through homodyne detection is optimized for every set of parameters and for all the figures of merit considered. We notice that the behaviours of these different optimized homodyne phases for squeezing, number of phonons and purity as a function of the detuning parameter are almost identical in all the cases we investigated.
In Fig. \ref{f:cavitymeasP} we plot the steady-state average phonon number, purity and squeezing for a reasonable  choice of the parameters entering in the master equation (\ref{eq:ME}). More specifically we fix these parameters following the experimentally reasonable assumptions made in Ref. \cite{Pflanzer2012}, for a silica nanosphere with a radius of $200$nm.
Note that very good results are obtained for all the values of the detuning we are considering. This is really important from an experimental point of view as it strongly relaxes the requirement to be sideband resolved in order to cool the nanosphere motion, a condition that in fact is particularly difficult to meet for low frequency oscillators.
As one can notice, if we are interested in cooling the oscillator, both in terms of number of phonons and of purity of the quantum state, the optimal choice is obtained either in the red or blue sideband for $\Delta \approx \pm 4 \omega_m$. On the other hand, in order to obtain the largest value of quantum squeezing, one may choose a value  of the detuning around $\Delta=-2.5\omega_m$, obtaining a non-classical state that exhibits around $3$dB of squeezing.
\subsection{Adding time-continuous measurement of the oscillator position} 
\begin{figure}[t]
\begin{center}
\includegraphics[width=0.8\columnwidth]{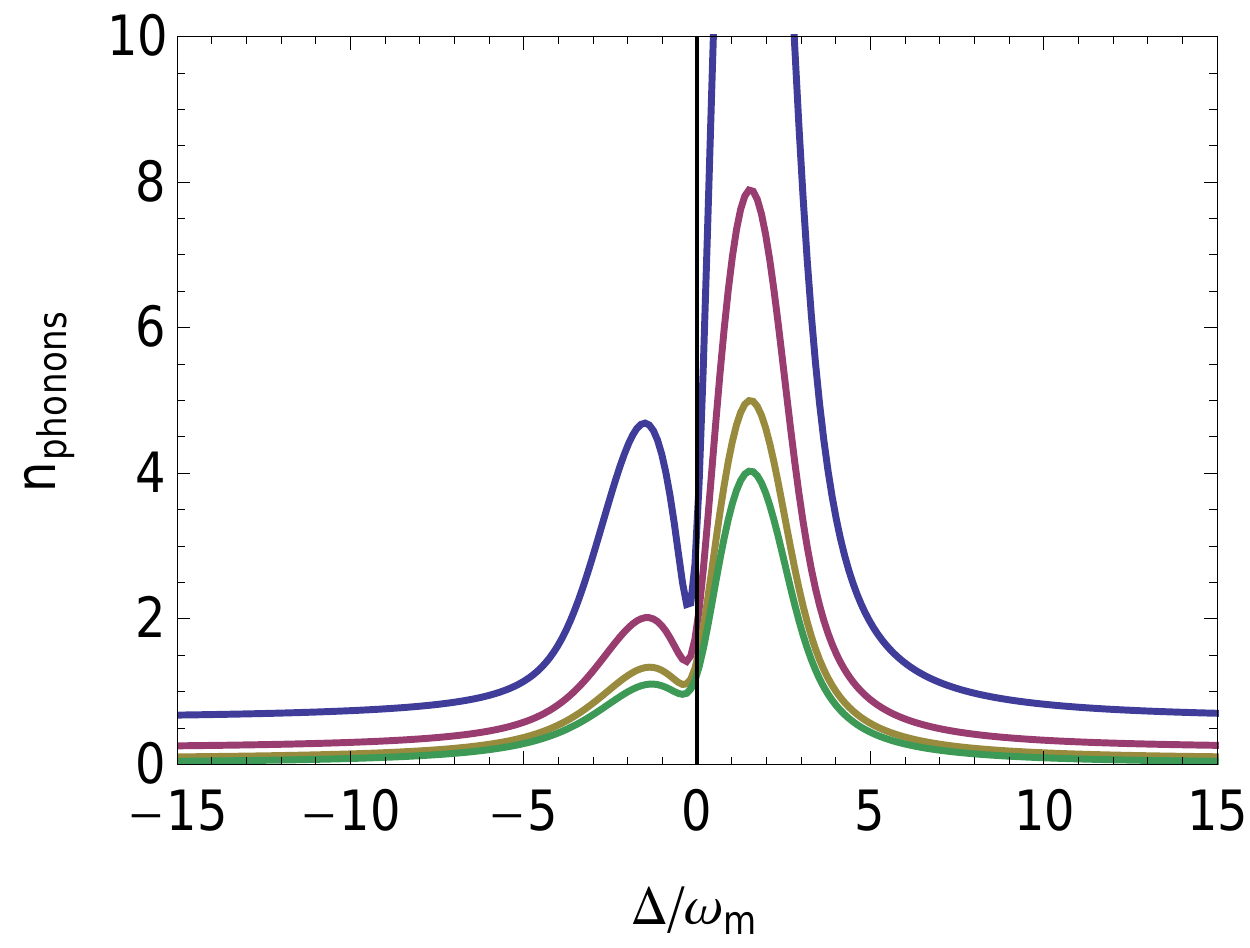}\ \\
\includegraphics[width=0.48\columnwidth]{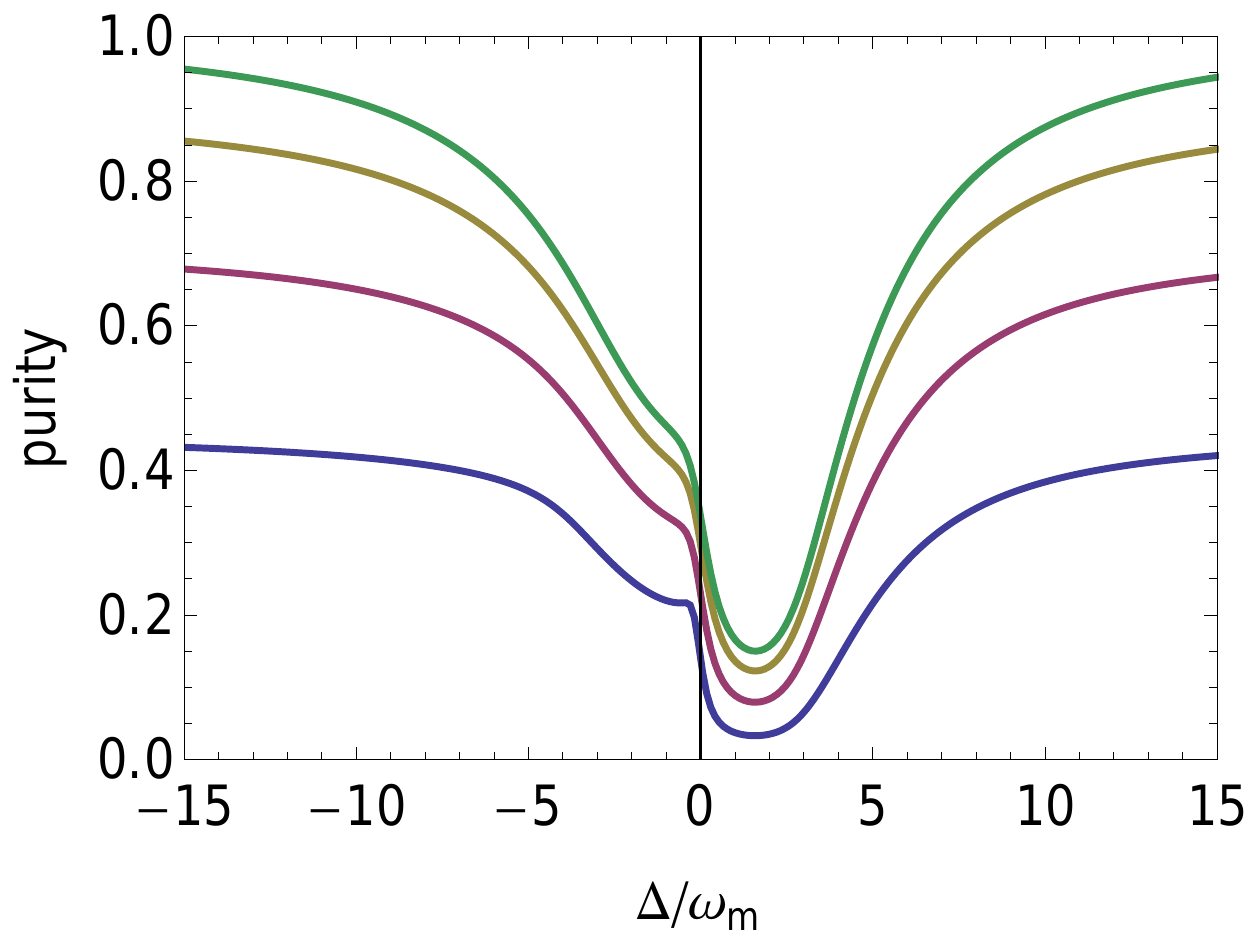} \medskip \
\includegraphics[width=0.48\columnwidth]{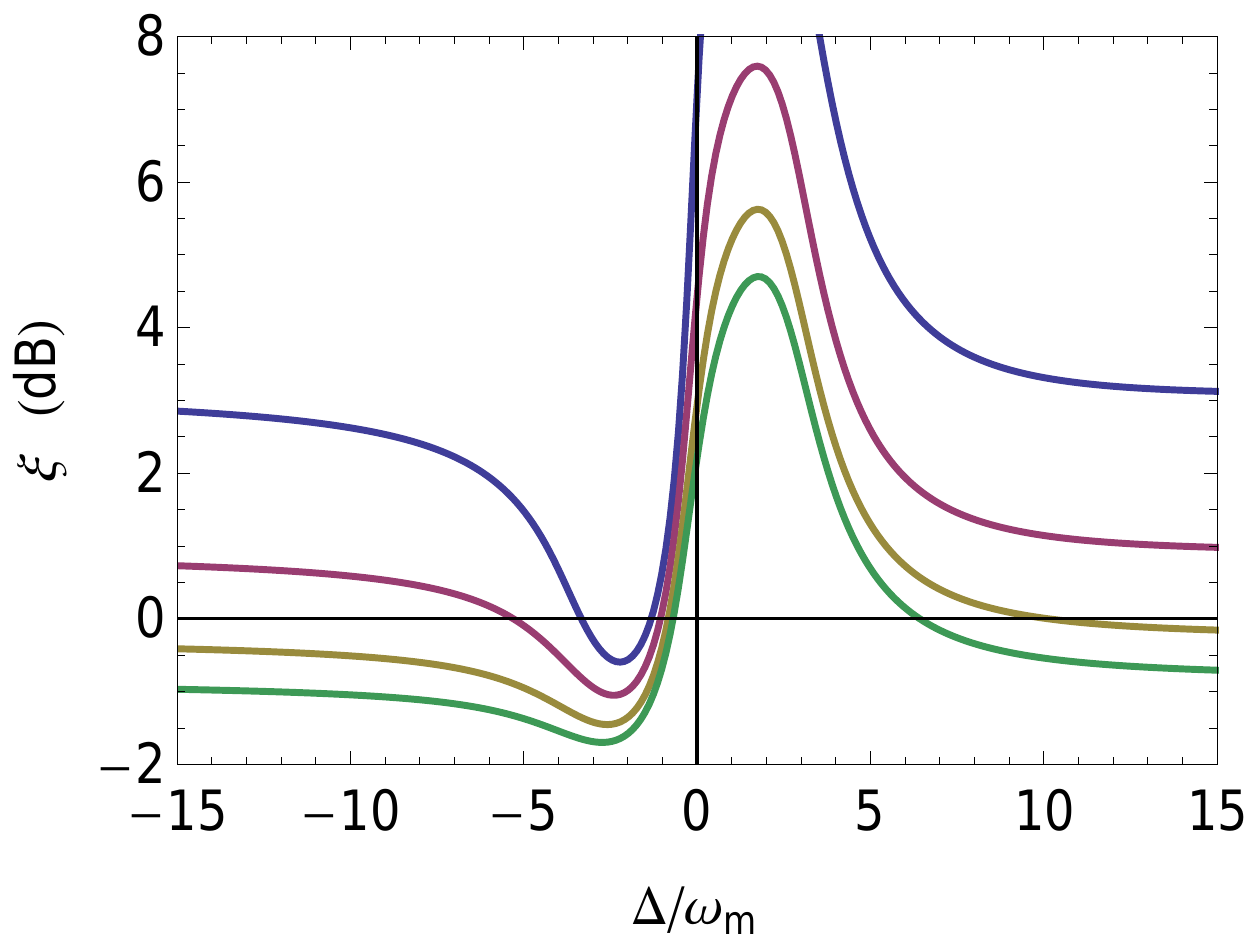}
\end{center}
\caption{Results obtained via monitoring the oscillator position. Top: Steady-state values for the average number of phonons. Bottom: steady-state oscillator purity (left) and quantum squeezing in dB scale (right). All quantities are plotted as a function of the detuning $\Delta$ and for different values of the oscillator position measurement efficiency: blue, $\eta_2=0.2$; purple, $\eta_2=0.5$; yellow, $\eta_2=0.8$; green $\eta_2=1$.
The other parameters are fixed as follows: $\eta_1=0$, $g=\omega_m$, $\kappa=2\omega_m$ and $\Gamma=\omega_m/10$.
\label{f:oscmeas}}
\end{figure}

\begin{figure}[t]
\begin{center}
\includegraphics[width=0.48\columnwidth]{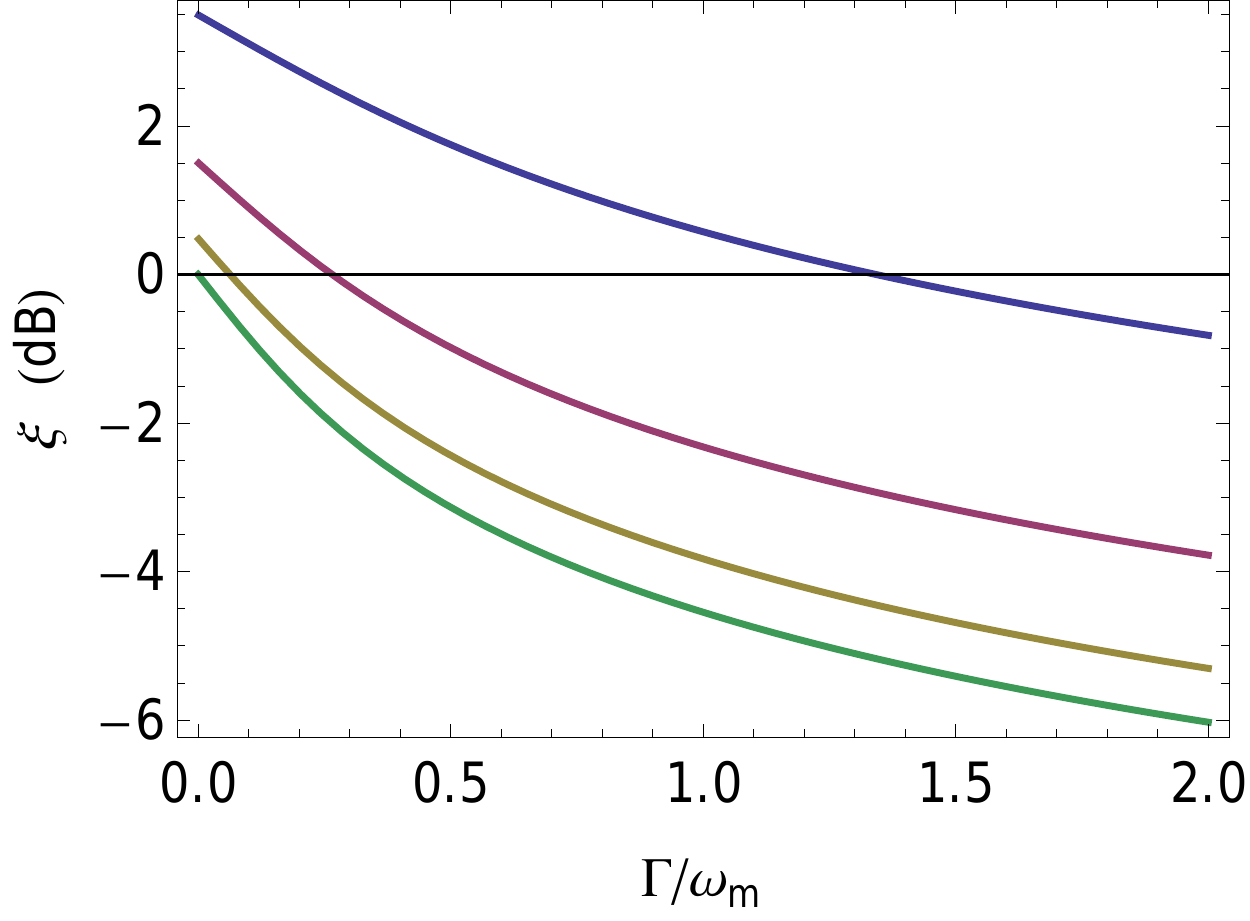} \medskip \
\includegraphics[width=0.48\columnwidth]{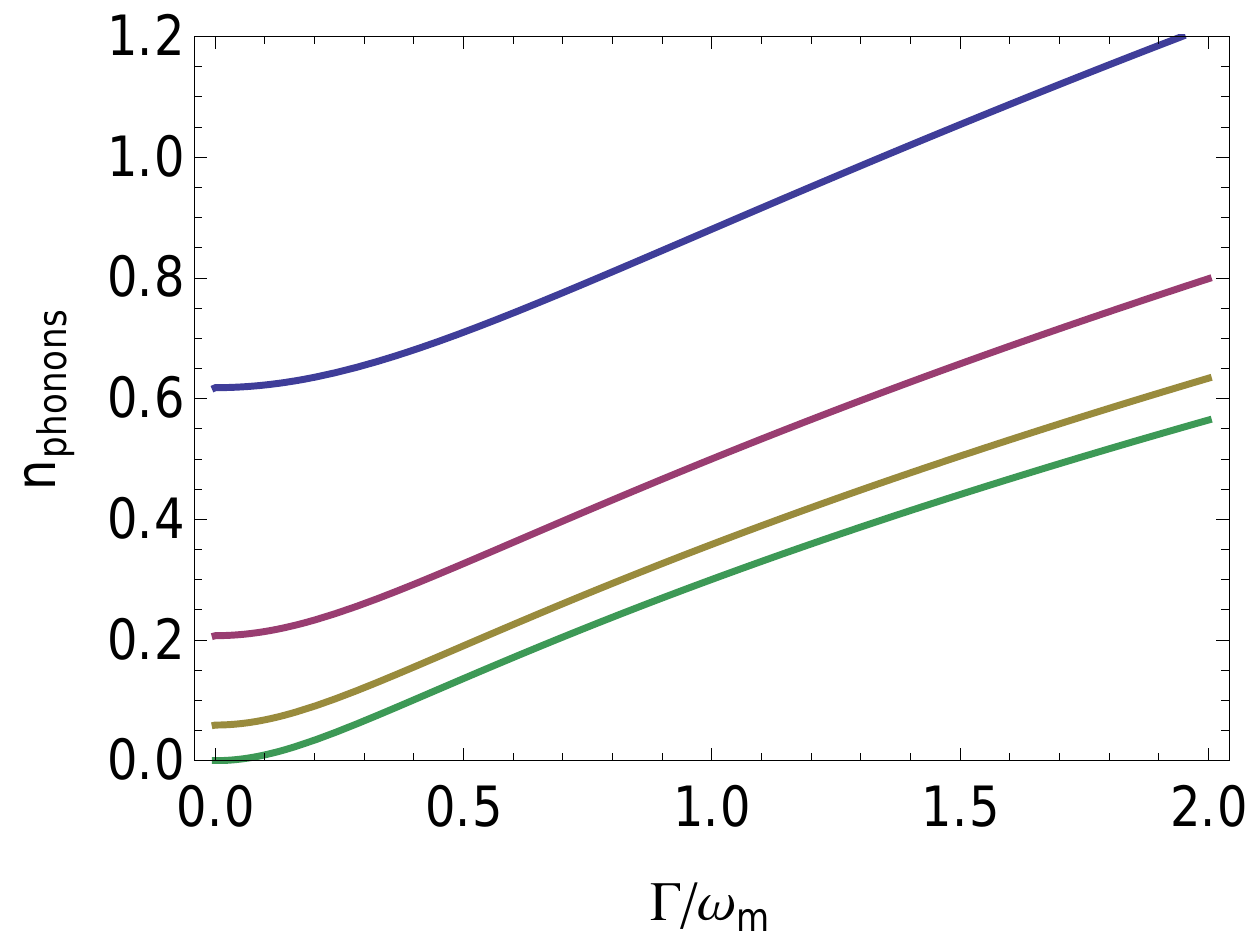}
\end{center}
\caption{
Results obtained through analytical calculations for steady-state squeezing in dB scale (left) and number of phonons (right) of a decoupled mechanical oscillator ($g=0$) subjected to time-continuous position measurement, as a function of the decoherence parameter $\Gamma$ and for different values of the measurement efficiency $\eta_2$: from top to bottom, $\eta_2=\{ 0.2, 0.5, 0.8, 1 \}$. \label{f:decoupled}}
\end{figure}

As a preliminary analysis, let us consider the effect of continuous monitoring of the oscillator position, whilst the cavity output 
is left unobserved ($\eta_1=0$). The results for this case are shown in Fig. \ref{f:oscmeas}.
We observe that, if we want to minimize the average number of phonons or maximize the purity, the optimal performances are obtained in the case of large detuning $|\Delta| \gg 1$, regardless of the red or blue shift of the driving field. 
This should not come as a surprise, as a large detuning corresponds 
to decoupling the oscillator from the cavity, and hence directly measuring an isolated degree of freedom.
By solving the dynamics for the decoupled mechanical oscillator alone (i.e., for $g=0$), one can evaluate analytically the corresponding steady-state. One may prove that its purity simply depends on the measurement efficiency, as $\mu=\sqrt{\eta_2}$, which thus univocally characterize the entropy of the steady-state; on the other hand quantum squeezing and number of phonons do depend also on the ratio between the noise parameter and the mechanical frequency $\Gamma/\omega_m$,  and their behaviour is plotted in Fig. \ref{f:decoupled} for different values of the measurement efficiency. 
We remind the reader that larger values of $\Gamma$ correspond to a large amount of scattered light from the nanosphere. On the one hand, this implies a larger amount of incoherent energy acquired by the oscillator due to the recoil heating process; on the other hand, it also corresponds to a large amount of information available for the continuous position measurement, and thus to the possibility to {\em convert} such incoherent energy into stead-state quantum squeezing. This explain the different behaviour we observe in Fig. \ref{f:decoupled} for quantum squeezing and number of phonons.

As anticipated above, the results derived here for the decoupled oscillator almost perfectly correspond to the ones reported in Fig. \ref{f:oscmeas} in the case of large detuning. Note that the small values of the number of phonons away from resonance are indeed due to the fact that the oscillator decoherence rate $\Gamma$ is relatively small with respect to $\omega_m$ (in the range of $\omega_m/10$).
For example, for a unit efficiency measurement ($\eta_2=1$), an almost pure quantum state is obtained with around $n_{\sf ph}=0.02$ phonons. The state is also squeezed, with a squeezing around $1$dB (a value compatible with the number of phonons obtained).
Our findings show that, if direct position monitoring with high efficiency were possible, feedback cooling would greatly outperform sideband cooling of the oscillator.
However, in practice, a decoupled cavity is not likely to be a favourable condition to work in, as the actual efficiencies of position measurements through scattered light are bound to be severely limited by a number of practical factors (one among all, the geometric impossibility of probing the whole solid angle of scattering).
Quite interestingly, 
at lower values of the measurement efficiency (i.e. $\eta_2 < 0.2$) or if our aim is to optimize the squeezing of the steady-state (also with larger values of $\eta_2$), a combination of sideband cooling and position measurements still yields the best results. More specifically, as regards quantum squeezing, the optimal detuning is again around $\Delta\approx-2.5\omega_m$, obtaining a quantum squeezing around $1$dB and $2$dB.
Note that, in this section, we are not taking into account the fact that, in principle, varying the detuning will change the number of photons inside the cavity, and thus also the effective coupling constant $g$  (such that larger values of detuning correspond to lower values of $g$). This effect will be properly taken into account in Sec.~\ref{s:bead}. 
%

\begin{figure}[t]
\begin{center}
\includegraphics[width=0.8\columnwidth]{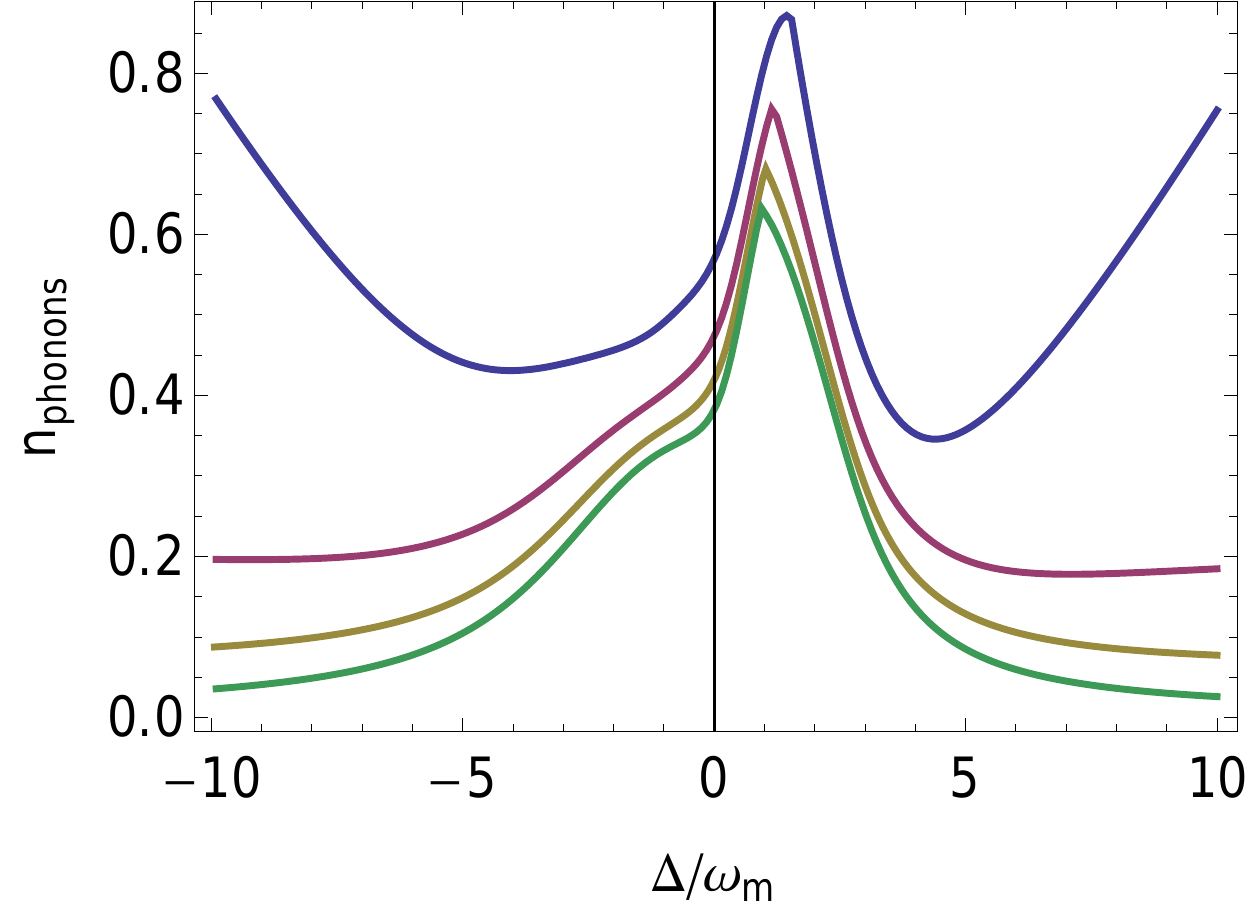}\ \\
\includegraphics[width=0.48\columnwidth]{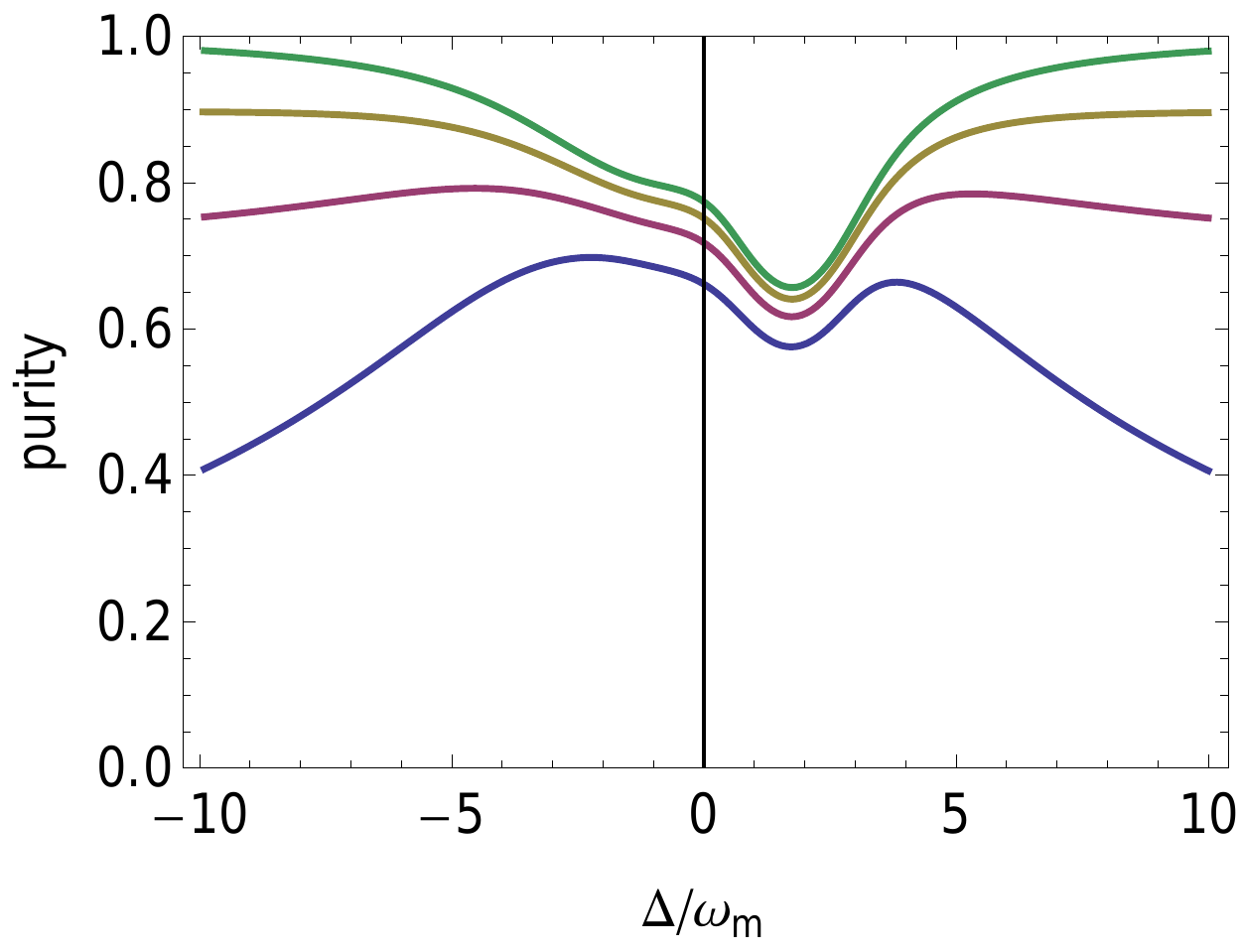} \medskip \
\includegraphics[width=0.48\columnwidth]{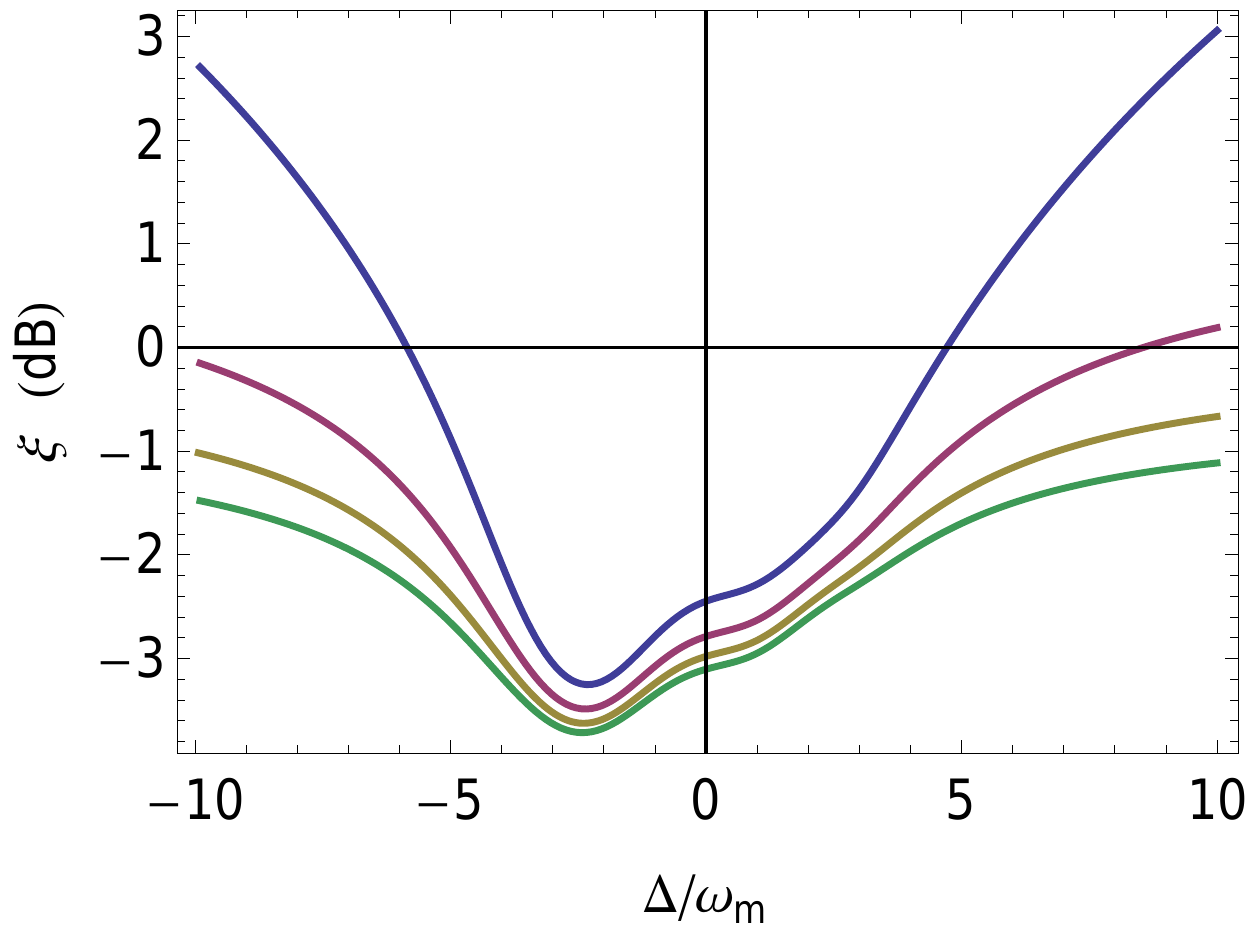}
\end{center}
\caption{Results for simultaneous optimized continuous homodyne of the cavity output mode (with unit efficiency) and of the oscillator position $x_m$. Top: Steady-state values for the average number of phonons. Bottom: steady-state oscillator purity (left) and quantum squeezing in dB scale (right). All quantities are plotted as a function of the detuning $\Delta$ and for different values of the oscillator position measurement efficiency: blue, $\eta_2=0$; purple, $\eta_2=0.5$; yellow, $\eta_2=0.8$; green $\eta_2=1$.
The other parameters are set as follows: $\eta_1=1$, $g=\omega_m$, $\kappa=2\omega_m$ and $\Gamma=\omega_m/10$.
\label{f:cavoscmeas}}
\end{figure}

Finally, let us consider the simultaneous monitoring of the cavity output mode (optimizing the phase of the homodyne detection and with unit efficiency: $\eta_1=1$) and of the oscillator position (with different efficiencies $\eta_2$). 
The results for this case are plotted in Fig. \ref{f:cavoscmeas}. 
The properties of the steady-state are qualitatively similar to the ones we have just discussed, showing the prominent role played by the oscillator position measurement over the other control strategies (i.e. sideband cooling and cavity homodyne measurement). Nevertheless, one observes slightly better results with respect to the unobserved cavity scenario, both in terms of steady-state phonons (in particular if we do not consider the large detuning regime) and in terms of quantum squeezing. For red-detuning with $\Delta\approx -2.5\omega_m$, one obtains the highest value of squeezing of approximately $4$dB.
\section{Results for a nanopshere levitated in a high finesse optical cavity} \label{s:bead}
 
\begin{figure}[t]
\begin{center}
\includegraphics[width=1\columnwidth]{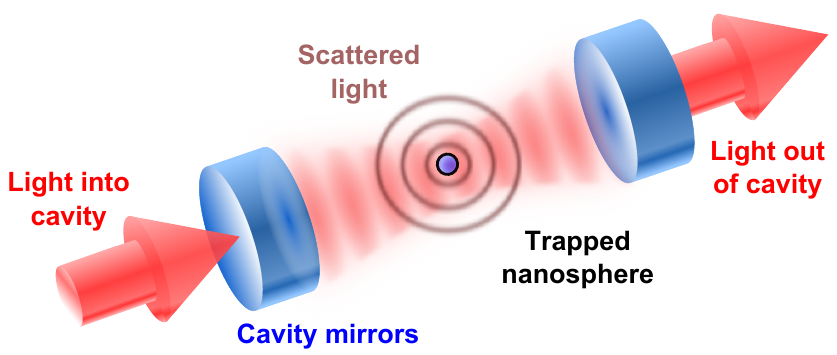}
\end{center}
\caption{Schematic of the experimental set-up for the quantum control of a levitated dielectric nanosphere within an optical cavity. The nanoparticle can be sideband-cooled by sending in light that is red-detuned from the cavity resonance. The light that leaves the cavity can be continuously monitored to perform generaldyne measurements. The light which is scattered by the particle can be collected to give information upon the position of the nanosphere.
\label{f:exp}}
\end{figure}

In this section, we will make specific predictions on what could be achieved by the continuous measurement of a nanosphere levitated by the field of a high finesse optical cavity, as depicted in Fig. \ref{f:exp}. The position and dynamics of the nanosphere can be directly measured by collecting the light it scatters or indirectly monitored through the homodyne monitoring of the light that leaves the optical cavity. A detailed description of this setup, comprising the derivation of the formulas we will use in the following can be found in \cite{Millen2014}.

The experimental setup we consider comprises a silica nanosphere of radius $r=200$ nm, with mass $m=7.35\cdot10^{-17}$Kg, and a cavity, characterized by a resonance frequency $\omega_c / 2\pi = 2.8 \cdot 10^{14}$ Hz ($\lambda=1064$ nm), length $L=13$ mm, finesse ${\cal F}=400000$, and cavity waist $w=60$ $\mu$m .
The corresponding value for the intrinsic cavity loss parameter is $\kappa_0/(2\pi)=c/(2 {\cal F} L)=29$ kHz. 
Following the results in \cite{Pflanzer2012}, we estimate that the extra loss due to the presence of the nanosphere inside the cavity is of the same order, $\kappa_d \sim \kappa_0$, such that the total loss parameter is fixed to $\kappa = 58$ kHz. The cavity average photon number reads
\begin{align}
n_c = \frac{\kappa}{2} \frac{P_{in}}{2\hbar (\Delta + \omega_c)} \frac{1}{\kappa^2 /4 + \Delta^2} .
\end{align}
As usual in opto-mechanical setups, the cavity-oscillator coupling constant $g$ depends on the cavity average photon number $n_c = |\alpha|^2$. In particular we adopt the formula
\begin{align}
g^2 \approx \frac{\hbar k^2 A^2}{2 m \omega_m} n_c,
\end{align}
where $k=2\pi/\lambda$, 
\begin{align}
A= \frac{3 V_s}{2 V_m} \frac{\epsilon_r -1}{\epsilon_r+2}\omega_l \:,
\end{align}
and the volumes of the nanosphere and of the cavity read $V_s =(4/3)\pi r^3$ and $V_m=\pi w^2 L$ respectively. The remaining parameters, $\epsilon_r$  and $\omega_l$ denote the dielectric constant and the driving laser frequency.
Since the nanosphere is here trapped by the cavity field, the mechanical frequency depends on the average photon number $n_c$ too, as per
\begin{align}
\omega_m^2 \approx \frac{2 \hbar k^2 A}{m} n_c .
\end{align}
It is important to notice that, because of the dependence of $\omega_m$ on the average photon number $n_c$,  the behaviour of the opto-mechanical coupling constant is modified with respect to standard setups, in that $g \sim n_c^{1/4}$. 

As far as the oscillator's decoherence rate $\Gamma$ is concerned, one can follow the results shown in \cite{Monteiro2013}. One observes that $\Gamma \sim n_c^{1/2}$, while the corresponding ratio $\Gamma/\omega_m \approx 0.15$ -- evaluated for our specific experimental parameters -- is fixed for every value of the detuning $\Delta$ \cite{note1}.
In the case of zero detuning ($\Delta=0$), we obtain a mechanical frequency $\omega_{m0}/2\pi \approx 33$ kHz and $g_0/2\pi \approx 20$ kHz.

Like in the previous section, we will consider the steady-state properties as a function of the detuning $\Delta$. As the mechanical frequency $\omega_m$ decreases for increasing detuning, the corresponding zero-point motion will increase; for this reason besides the quantum squeezing property (which is obtained by considering variances renormalized to the zero-point motion and thus does not take into account the effect of the variation of the frequency $\omega_m$), we will also consider the proper fluctuation of the position 
\begin{align}
\delta X = \sqrt{\frac{\hbar\:  \delta x_m^2}{m \omega_m} }\:,
\end{align}
$\delta x_m^2$ being the fluctuations of the dimensionless position operator $x_m$.  Apart from taking into account the different zero-point fluctuations, $\delta X$ is arguably more interesting from an experimental and practical point of view as it is directly accessible in experiments.

\begin{figure}[t!]
\begin{center}
\includegraphics[width=0.8\columnwidth]{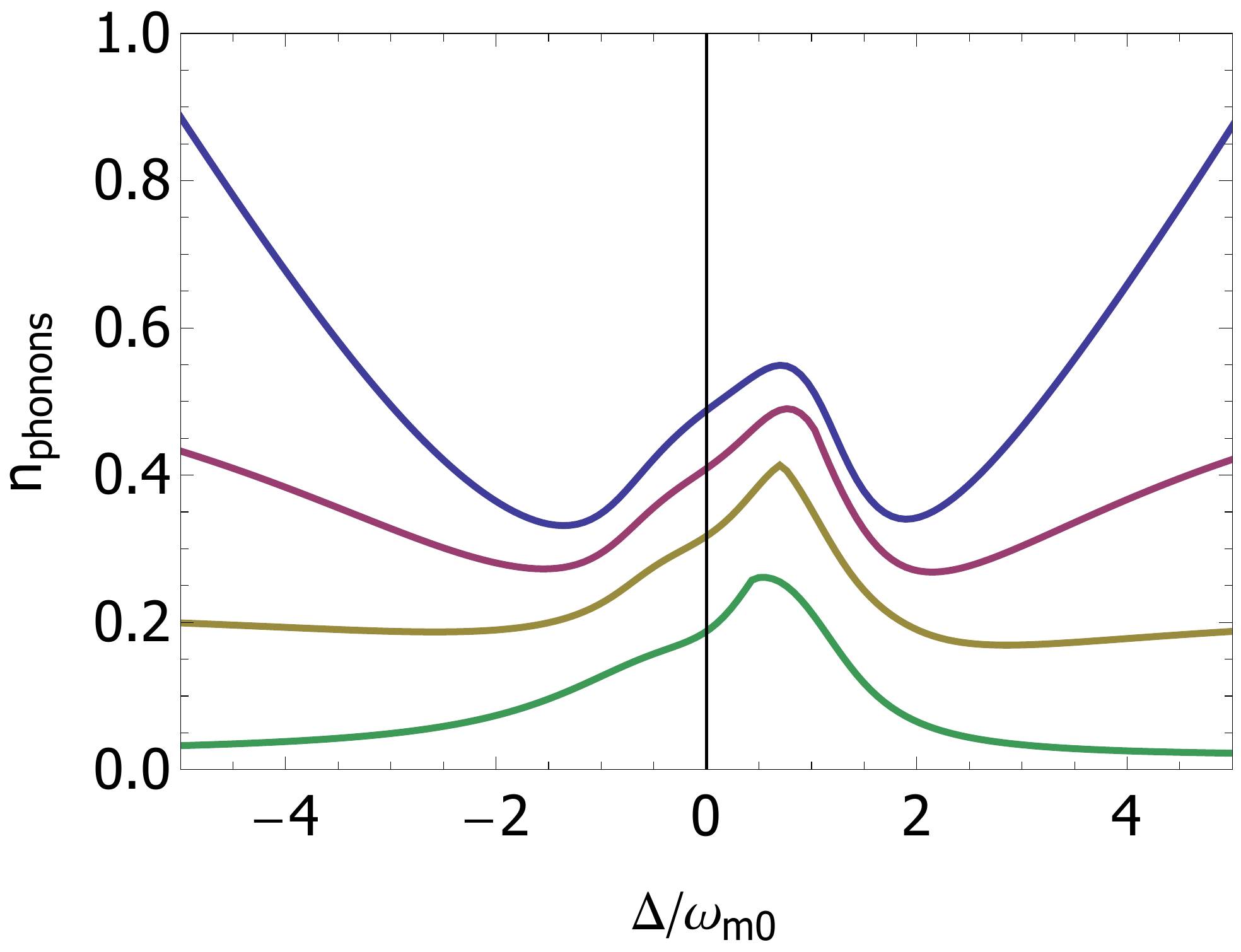}\ \\
\includegraphics[width=0.47\columnwidth]{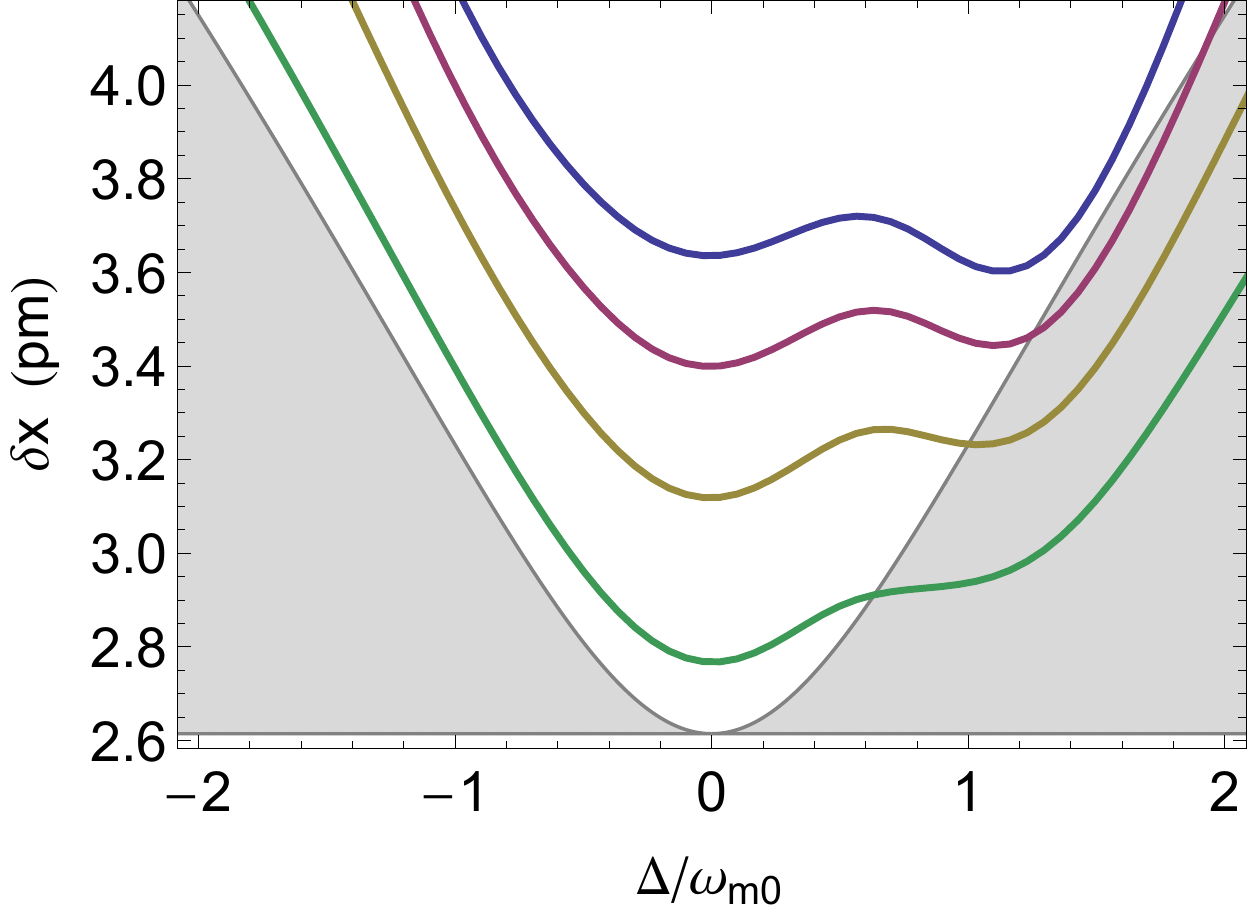} \:\: \
\includegraphics[width=0.48\columnwidth]{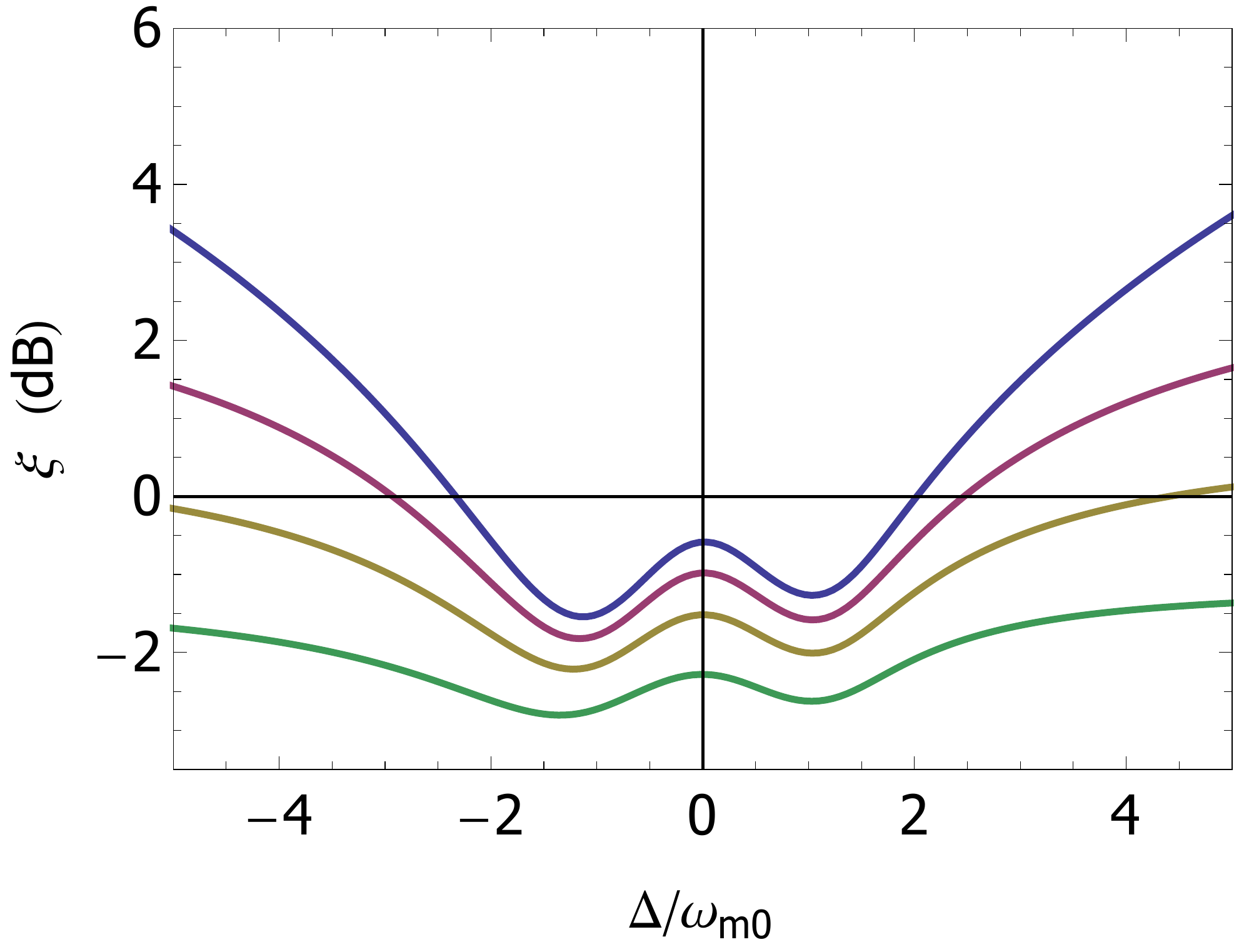}
\end{center}
\caption{
Results for simultaneous optimized homodyne monitoring of the cavity output mode (with unit efficiency) and of the oscillator position $x_m$. Top: steady-state value for the average phonon number of the oscillator. Bottom left: steady-state position uncertainty $\delta X$ (solid lines), with the gray area corresponding to sub vacuum fluctuations. Bottom right: steady-state quantum squeezing in dB scale.  
All quantities are plotted as a function of the detuning $\delta$ (in units of the mechanical frequency $\omega_{m0}$, and for different values of the oscillator position measurement efficiency: blue, $\eta_2=0$; purple, $\eta_2=0.2$; yellow, $\eta_2=0.5$; green $\eta_2=1$.
The other experimental parameters are fixed as described in the main text.
\label{f:nanosphereOPT}}
\end{figure}

\begin{figure}[t!]
\begin{center}
\includegraphics[width=0.8\columnwidth]{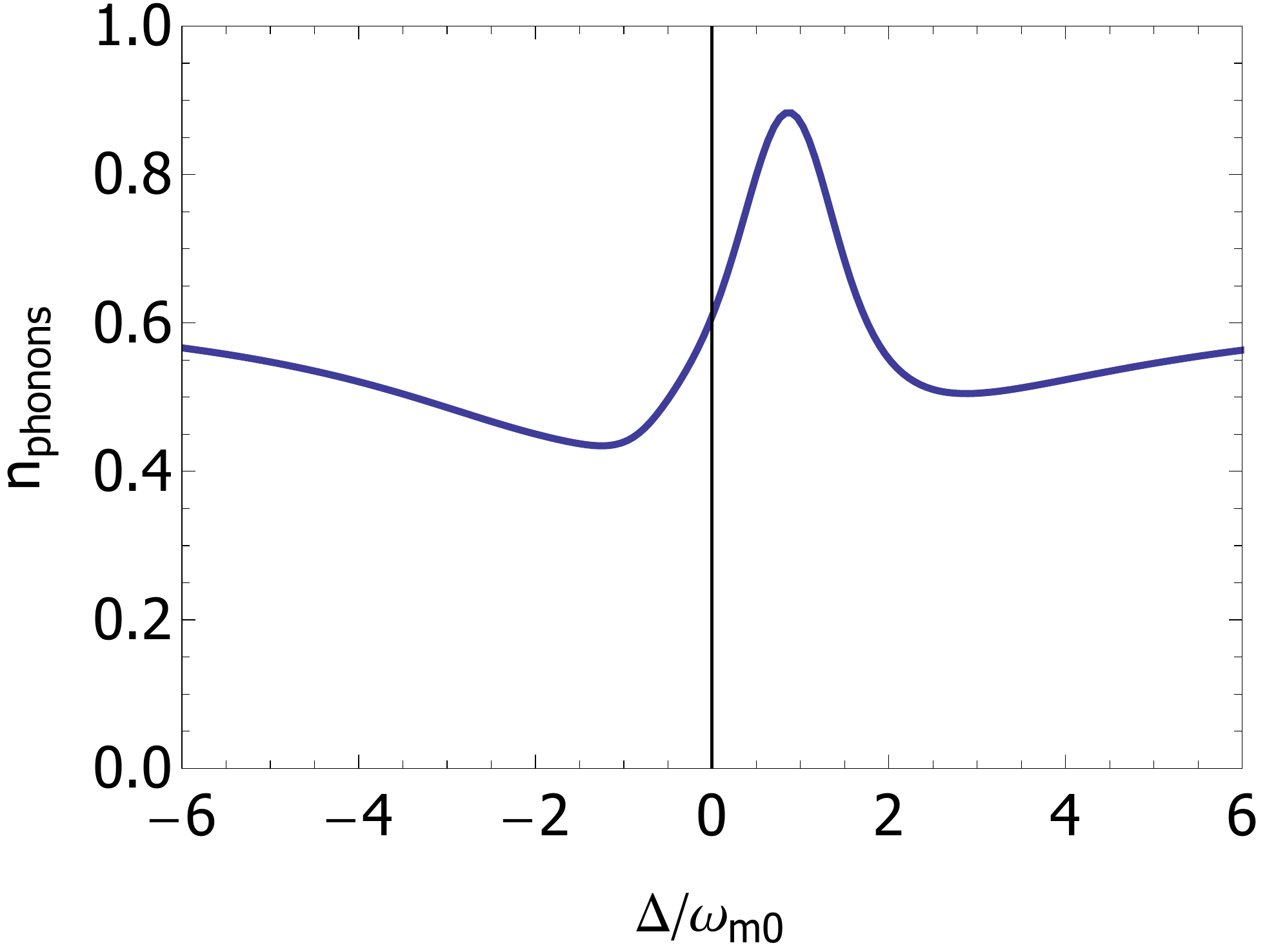} \ \\
\includegraphics[width=0.47\columnwidth]{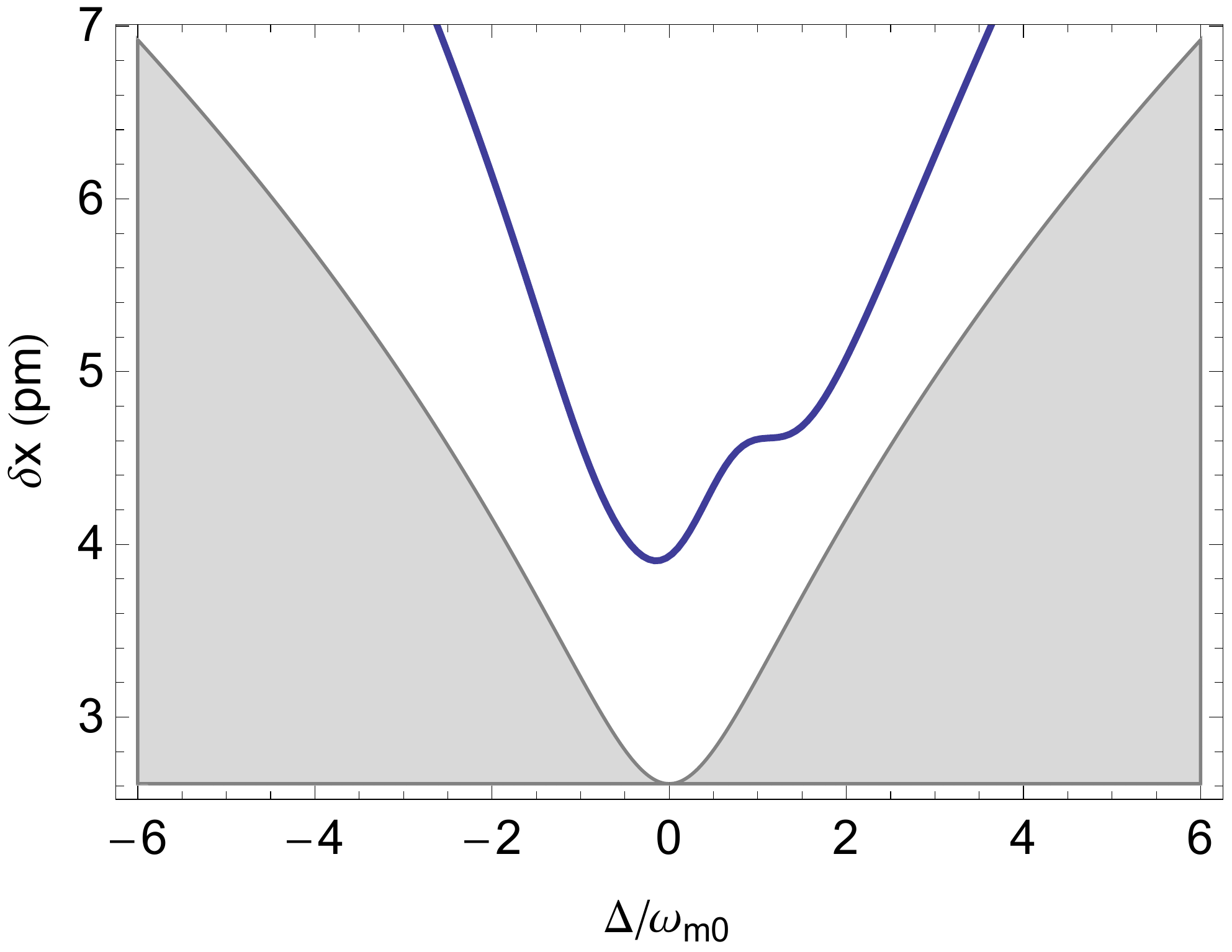} \: \: 
\includegraphics[width=0.47\columnwidth]{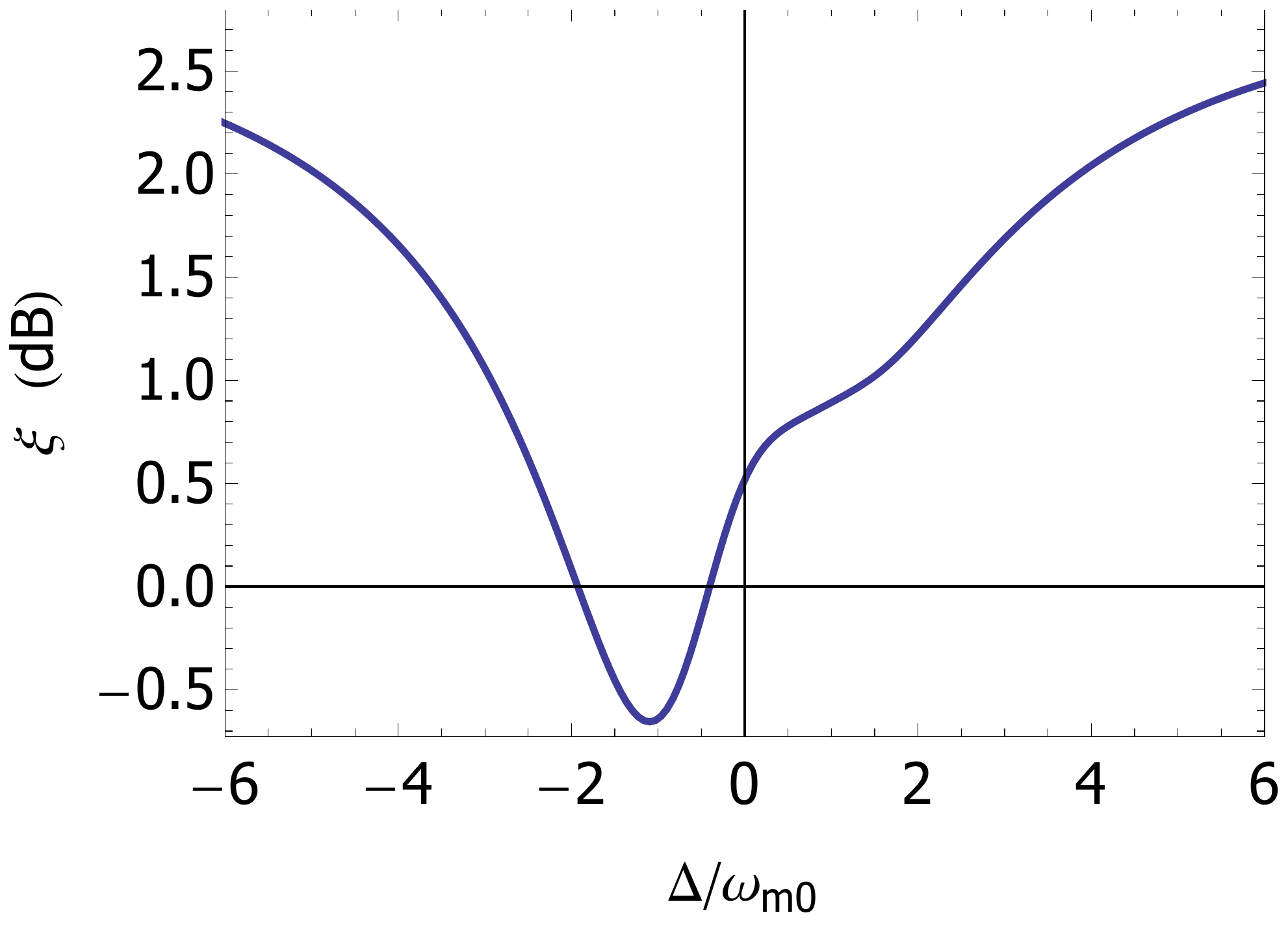}
\end{center}
\caption{
Results for simultaneous optimized homodyne monitoring of the cavity output mode and of the oscillator position $x_m$, with realistic values of the measurement efficiencies: $\eta_1=0.5$ and $\eta_2=0.2$. 
Top: steady-state value for the average phonon number of the oscillator. Bottom left: steady-state position uncertainty $\delta X$ (solid lines), with the gray area corresponding to sub vacuum fluctuations. Bottom right: steady-state quantum squeezing in dB scale.
All quantities are plotted as a function of the detuning $\Delta$ (in units of the mechanical frequency $\omega_{m0}$). The other experimental parameters are fixed as described in the main text.
\label{f:nanosphereREAL}}
\end{figure}

The results are depicted in Figs. \ref{f:nanosphereOPT} and \ref{f:nanosphereREAL}. In the first one, we plot the results in the case where the efficiency of the homodyne measurement of the cavity output is maximum ($\eta_1=1$), the phase of the homodyne is optimized for each figure of merit, and different values of the oscillator's measurement efficiency $\eta_2$ are considered. As one should expect, we observe better performances for increasing values of $\eta_2$. 
In detail, by focusing on the average number of phonons, for small values of $\eta_2$ we observe that two local minima of $n_{\sf ph}$ are obtained, one in the red- and one in the blue-detuning regimes. 
By considering an efficiency over $50\%$, the optimal performances are obtained in the case of large detuning where, as discussed above, the oscillator's motion decouples from the cavity mode, and thus its steady-state properties become entirely dependent on the position measurement efficiency. If we instead focus on quantum squeezing, in this case too we obtain that for all values of $\eta_2$ two local minima are present in the red and blue sideband, and the optimal working point corresponds to a value of the detuning near $\Delta = - 3\omega_{m0}/2$. Surprisingly, observing the behaviour of the position fluctuations $\delta X$, its actual minimum is always reached at resonance, while sub-vacuum fluctuations are observed only in the blue sideband, that is for $\Delta >0$.
One important remark is needed here: as sub-vacuum fluctuations are not obtained for most of the values of the detuning $\Delta$, the amount of quantum squeezing observed in the other plot of Fig. \ref{f:nanosphereOPT} necessarily corresponds to different quadrature operator describing the oscillator; in general, this quadrature will correspond to a certain linear combination of position and momentum of the oscillator, whose usefulness and detectability may not be straightforward. 

Similar observations are made also regarding Fig. \ref{f:nanosphereREAL}, where we considered plausible values for the position and homodyne measurement efficiencies, estimated respectively at $\eta_2=0.2$, based on the the set-up of Ref. \cite{Millen2014}, and at $\eta_1=0.5$ corresponding to the homodyne set-up efficiency in \cite{Safavi2013}. In this realistic scenario, we demonstrate that for all the different values of $\Delta$ considered one can achieve a steady-state characterized by $n_{\sf ph}<1$, with a minimum reaching $n_{\sf ph}\approx 0.4$ and a maximum squeezing of $\hbox{dB}\approx -0.5$ both obtained in the red sideband. 
\section{Summary and conclusions} \label{s:conclusions}
Our study explicitly shows the quantum control possibilities offered, in realistic setups, by the combined simultaneous monitoring
of scattered as well as coherent cavity light interacting with a levitating dielectric nanosphere.
In particular, it was shown that
\begin{itemize}
\item{time-continuous measurements of either the cavity mode or the oscillator position, accompanied by Markovian feedback, are able to stabilize the nanosphere motion for all the values of the detuning $\Delta$ and of the measurements efficiencies.}
\item{The addition of time-continuous homodyne monitoring of the cavity output plus Markovian feedback greatly improves the performance that one would have obtained with sideband cooling only. 
For the realistic values of physical parameters considered in our study, while sideband cooling would prepare a phase-insensitive steady-state characterized by $n_{\sf ph} \gtrsim 10$ phonons on average, the addition of continuous homodyne measurements of the cavity output would prepare a squeezed steady-state, with $n_{\sf ph} < 2$ phonons. In particular, this is true for a large range of detuning values, which relaxes the requirements of sideband resolution to cool down the oscillator. This is particularly important for low frequency heavy oscillators which cannot be operated in the sideband resolved regime for dispersive coupling or where a dissipative coupling is not available
\cite{disscoupling}.}
\item{
In terms of optimizing the purity and the cooling of the oscillator, sideband cooling would cease to be useful if one is able to directly measure the oscillator with very high efficiency. 
Nevertheless, if we take into account the state-of-the-art values for these measurement efficiencies in an actual experimental setup, the combination of the two control procedures represents still the best choice for experimental realizations, leading in principle to a quantum squeezed steady-state with less than one phonon on average. The performances are further improved if a simultaneous measurement of the cavity mode output is carried out.}
\item{
If we consider a state-of-the-art experimental setup where the dielectric nanosphere is trapped by the field of a high-finesse cavity, the proposed measurement protocols are in principle able to prepare a quantum squeezed state with less than $n_{\sf ph} =1$ phonons at steady-state.}
\end{itemize}
Over the next few years, it will be possible to perform more and more exhaustive time-continuous measurements 
on the outputs of interesting micro- and meso-scopic physical systems \cite{Kalman}.
It is apparent from our findings that such a possibility 
will be one of the pathways to reduce the entropic content of such systems, drive them to the quantum regime,
and ultimately achieve their full or partial quantum control.\\

{\em Note added}: after this work was completed, a similar approach was presented in \cite{Rodenburg2015} for a different experimental setup in which the nanosphere is trapped without the help of a optical cavity.

\section{Acknowledgements}
MGG, JM, PFB and AS acknowledge support from EPSRC through grant EP/K026267/1.
\appendix
\section{Unconditional and conditional evolution for first and second moments} \label{s:appendix}
In this appendix, we provide the explicit form of the matrices entering the unconditional and conditional evolution equations for first and second moments, corresponding to the Eqs. (\ref{eq:ME}) and (\ref{eq:SME}), derived following the general framework provided in \cite{WisemanDoherty}. 

As we stated in the manuscript, the master equation (\ref{eq:ME})) describing the unconditional noisy evolution of the quantum state of the mechanical mode and of the cavity mode, yields the following evolution for the corresponding first moment vector ${\bf R}$ and for the covariance matrix $\sigmaCM$:
\begin{align}
\frac{d {\bf R}}{dt} &= A {\bf R}\:,  \\
\frac{d \sigmaCM}{dt} &= A \sigmaCM + \sigmaCM A^{\sf T} + D \:.
\end{align}
The drift matrix $A$ and the diffusion matrix $D$ read
\begin{align}
A &=
\left(
\begin{array}{c c c c} 
-\frac{\kappa}{2} & - \Delta & 0 & 0 \\
\Delta & -\frac{\kappa}{2} & -2 g & 0 \\
0 & 0 & 0 & \omega_m \\
-2g & 0 & -\omega_m & 0
\end{array}
\right), \\
D &=
\left(
\begin{array}{c c c c} 
\kappa &0 & 0 & 0 \\
0 & \kappa & 0 & 0 \\
0 &0 & 0 & 0 \\
0 &0 & 0 & 4 \Gamma \\
\end{array}
\right) . 
\end{align}
On the other hand, if we consider simultaneous continuous monitoring of the output cavity mode with homodyne detection, and of the oscillator position, the corresponding stochastic master equation (\ref{eq:SME}) is translated for first moment vector and covariance matrix in:
\begin{align}
d {\bf R} &= A {\bf R} \:dt + ( N - \sigmaCM B^{\sf T}) \:d{\bf w} \\
\frac{d \sigmaCM}{dt} &= \widetilde{A} \sigmaCM + \sigmaCM {\widetilde{A}}^{\sf T} -\sigmaCM B^{\sf T} B \sigmaCM + \widetilde{D} \:, 
\end{align}
where $d{\bf w} = (dw_1, dw_2)^{\sf T}$, $A$ has been defined just above and the other matrices read
\begin{align}
B &=
\left(
\begin{array}{c c c c} 
\sqrt{\eta_1\kappa} \cos^2 \phi & - \sqrt{\eta_1\kappa} \sin\phi\cos\phi & 0 & 0 \\
-\sqrt{\eta_1\kappa} \sin\phi\cos \phi &\sqrt{\eta_1\kappa} \sin^2 \phi & 0 & 0 \\
0 & 0 & 0 & 0\\
0 & 0 & \sqrt{4 \eta_2 \Gamma}  & 0
\end{array}
\right) , \\
N &=
\left(
\begin{array}{c c c c} 
\sqrt{\eta_1\kappa} \cos^2 \phi & - \sqrt{\eta_1\kappa} \sin\phi\cos\phi & 0 & 0 \\
-\sqrt{\eta_1\kappa} \sin\phi\cos \phi &\sqrt{\eta_1\kappa} \sin^2 \phi & 0 & 0 \\
0 & 0 & 0 & 0 \\
0 & 0 & 0 & 0
\end{array}
\right)  , \\
\widetilde{A} &= A + N B, \\
\widetilde{D} &= D - N N^{\sf T} . 
\end{align}

\begin{thebibliography}{99}
\bibitem{note1} Notice that in principle one should also consider an additional phenomenological momentum damping due to collision with the surrounding gas particles. However one can show that, in the present experimental conditions with pressure around $10^{-9}$ bar,  the corresponding rate is negligible compared to the other dynamical parameters entering in the master equation (\ref{eq:ME})
\end{thebibliography}
\end{document}